\documentclass{rQUF2e}
\usepackage{epstopdf}% To incorporate .eps illustrations using PDFLaTeX, etc.
\usepackage{subfigure}% Support for small, `sub' figures and tables
\usepackage{tabulary}
\usepackage{longtable}
\usepackage{capt-of}
\usepackage{algpseudocode}
\usepackage{algorithm}
\usepackage{tikz}
\usepackage{amsmath}
\usepackage[unicode=true,pdfusetitle,
bookmarks=true,bookmarksnumbered=false,bookmarksopen=false,
breaklinks=false,pdfborder={0 0 0},backref=false,colorlinks=false]{hyperref}
\usepackage[noabbrev]{cleveref}
\usepackage[shortlabels]{enumitem}
\usepackage{makecell}
\usepackage{array}
\usepackage[labelsep=none]{caption}
\DeclareCaptionLabelFormat{mylabel}{#1 #2. \hspace{1ex}}
\newcommand{\Rho}{\mathrm{P}}
\DeclareMathOperator{\EX}{\mathbb{E}}% expected value
\newcommand{\be}{\begin{equation}}
\newcommand{\ee}{\end{equation}}

\theoremstyle{plain}
\newtheorem{theorem}{Theorem}[section]

\newtheorem{proposition}[theorem]{Proposition}

\theoremstyle{definition}
\newtheorem{definition}{Definition}

\theoremstyle{remark}

\begin{document}

%\jvol{00} \jnum{00} \jyear{2014} \jmonth{October}

\title{Learning the dynamics of technical trading strategies}

\author{N. J. MURPHY$^{\ast}$$\dag$\thanks{$^\ast$Corresponding author.} and T. J.  GEBBIE$^\S$${\dag}$\thanks{$^\S$
Email: tim.gebbie@uct.ac.za}\\
\affil{$\dag$ Department of Statistical Sciences, University of Cape Town, Rondebosch, Cape Town, South Africa}}

\maketitle

\begin{abstract}
We use an adversarial expert based online learning algorithm to learn the optimal parameters required to maximise wealth trading zero-cost portfolio strategies. The learning algorithm is used to determine the relative population dynamics of technical trading strategies that can survive historical back-testing as well as form an overall aggregated portfolio trading strategy from the set of underlying trading strategies implemented on daily and intraday Johannesburg Stock Exchange data. The resulting population time-series are investigated using unsupervised learning for dimensionality reduction and visualisation. A key contribution is that the overall aggregated trading strategies are tested for statistical arbitrage using a novel hypothesis test proposed by \cite{jarrow} on both daily sampled and intraday time-scales. The (low frequency) daily sampled strategies fail the arbitrage tests after costs, while the (high frequency) intraday sampled strategies are not falsified as statistical arbitrages after costs. The estimates of trading strategy success, cost of trading and slippage are considered along with an offline benchmark portfolio algorithm for performance comparison. In addition, the algorithms generalisation error is analysed by recovering a probability of back-test overfitting estimate using a nonparametric procedure introduced by \cite{pbo}. The work aims to explore and better understand the interplay between different technical trading strategies from a data-informed perspective.
\end{abstract}


\begin{classcode}G11, G14 and O55\end{classcode}
 
\section{Introduction}

Maximising wealth concurrently over multiple time periods is a difficult task; particularly when combined with capital allocations that are made simultaneously with the selection of plausible candidate trading strategies and signals. An approach to combining strategy selection with wealth maximisation is to use online or sequential machine learning algorithms (\cite{gyorfibook}). Online portfolio selection algorithms attempt to automate a sequence of trading decisions among a set of stocks with the goal of maximising returns in the long run. Here the long run can correspond to months or even years, and is dependent on the frequency at which trading takes place.\footnote{This could imply anything from a few days to a few weeks for high frequency trading algorithms, and a few months (years) for daily (weekly) trading algorithms.} Such algorithms typically use historical market data to determine, at the beginning of a trading period, a way to distribute their current wealth among a set of stocks. These types of algorithms can use many more features than merely prices, so called ``side-information'', but the principle remains that same. 

The attraction of this approach is that the investor does not need to have any knowledge about the underlying distributions that could be generating the stock prices (or even if they exist). The investor is left to ``learn'' the optimal portfolio to achieve maximum wealth using past data directly (\cite{gyorfibook}).  

\cite{cover} introduced a ``follow-the-winner" online investment algorithm\footnote{{\it Follow-the-winner} algorithms give greater weightings to better performing experts or stocks (\cite{binli})} called the Universal Portfolio (UP) algorithm\footnote{The algorithm was later refined by \cite{coverOrd} (see \cref{sec:onlinelearn})}. The basic idea of the UP algorithm is to allocate capital to a set of experts characterised by different portfolios or trading strategies; and to then let them run while at each iterative step to shift capital from losers to winners to find a final aggregate wealth.

Here our ``experts'' will be similarly characterised by a portfolio, but portfolios that proxies different trading strategies. Here a particular agent makes decisions independently of all other experts. The UP algorithm holds parametrised Constant Rebalanced Portfolio (CRP) strategies as its underlying experts. We will have a more generalised approach to generating experts. The algorithm provides a method to effectively distribute wealth among all the CRP experts such that the average log-performance of the strategy approaches the Best Constant Rebalanced Portfolio (BCRP) which is the hindsight strategy chosen which gives the maximum return of all such strategies in the long run. A key innovation was the provision of a mathematical proof for this claim based on arbitrary sequences of ergodic and stationary stock return vectors \cite{cover}. 

If some log-optimal portfolio exists such that no other investment strategy has a greater asymptotic average growth then to achieve this one must have full knowledge of the underlying distribution and of the generating process to achieve such optimality (\cite{algoet,cover,coverOrd,gyorfi}). Such knowledge is unlikely in the context of financial markets. However, strategies which achieve an average growth rate which asymptotically approximates that of the log-optimal strategy are possible when the underlying asset return process is sufficiently close to being stationary and ergodic. Such a strategy is called \textit{universally consistent}. 

\cite{gyorfi} proposed a universally consistent portfolio strategy and provided empirical evidence of a strategy based on nearest-neighbour based experts which reflects such asymptotic log-optimality. The idea is to match current price dynamics with similar historical dynamics using pattern matching. The pattern matching is implemented using a nearest-neighbour search algorithm to select parameters for experts. The pattern-matching algorithm was extended by \cite{loonat} in order to implement a zero-cost portfolio {\it i.e.} a long/short and self-financing portfolio. The algorithm was also re-cast to replicate near-real-time applications using look-up libraries learnt offline. However, there is a computational cost associated with coupling and creating offline pattern libraries and the algorithms are not truly online. 

A key objective in the implementation of online learning here is that the underlying experts remain online and they can be sequentially computed on a moving finite data-window using parameters from the previous time-step. Here we have ignored the pattern-matching step in the aforementioned algorithms and rather propose our own expert generating algorithm using tools from technical analysis. Concretely, we replace the pattern-matching expert generating algorithm with a selection of technical trading strategies.

Technical analysis indicators are popular tools from technical analysis used to generate trading strategies (\cite{chan2009,clayburg}). They claim to be able to exploit statistically measurable short-term market opportunities in stock prices and volume by studying recurring patterns in historical market data (\cite{creamer,chande1994,PHDRechenthin}). What differentiates technical analysis from traditional time-series analysis is that it tends to place an emphasis on recurring time-series patterns rather than invariant statistical properties of times-series. Traditionally, technical analysis has been a visual activity, whereby traders study the patterns and trends in charts, based on price or volume data, and use these diagnostic tools in conjunction with a variety of qualitative market features and news flow to make trading decisions. 

This is perhaps not dissimilar to the relationship between alchemy and chemistry, or astrology and astronomy, but in the setting of financial markets, but many studies have criticised the lack of a solid mathematical foundation for many of the proposed technical analysis indicators (\cite{Aronson,lo2000,hereticsfinance}). There has also been an abundance of academic literature, utilising technical analysis for the purpose of trading, and several studies have attempted to develop indicators and test them in a more mathematically, statistically and numerically sound manner (\cite{Aronson,QuantStrat,techanreview}). Much of this work is still viewed with some suspicion given that it is extremely unlikely that this or that particular strategy or approach was not the result of some sort of back-test overfitting (\cite{phacking, bailey, pbo}). 

Our work does not address the question: ``Which, if any, technical analysis methods reveal useful information for trading purposes?''; rather, we aim to bag a collection of technical experts and allow them to compete in an adversarial manner using the online learning algorithm. This allows us to consider whether the resulting aggregate strategy can: 1.) pass reasonable tests for statistical arbitrage, and 2.) has a relatively low probability of being the result of back-test overfitting. Can the strategy be considered a statistical arbitrage, and can it generalise well out-of-sample? 

Concretely, here we are concerned with the idea of understanding whether the collective population of technical experts can through time lead to dynamics that can be reasonably be considered a statistical arbitrage (\cite{jarrow}), and then with a reasonably low probability of back-test overfitting (\cite{pbo}). Can we generate wealth, both before costs, and then after costs, using the online aggregation of technical strategies? Then, what broad groups of strategies will emerge as being successful in the sense of positive trading profits with declining variance in losses? 

Incorrectly accounting for costs will always be a plausible explanation for any apparently profitable trading strategy (see \cite{loonat}), but even after costs there still exists a high likelihood that there was some data-overfitting because we only have single price path from history with little or no knowledge about the true probability of the particular path that has been measured. While the adaptive nature of markets themselves are continually changing the efficacy of various strategies and approaches. 

Rather than considering various debates relating the technicalities of market efficiency, where one is concerned with the expeditiousness of market prices to incorporate new information at any time, and where information is explicitly exogenous; we restrict ourselves to market efficiency in the sense used by Fischer Black (\cite{black, bouchaudfactor2, Aronson}). This is the situation where some of the short-term information is in fact noise, and that this type of noise is a fundamental property of real markets. Although market efficiency may plausibly hold over the longer term, in the short-term there may be small departures that are amenable to tests for statistical arbitrage (\cite{jarrow}), departures that create incentives to trade, and more importantly, departures that may not be easily traded out of the market due to various asymmetries in costs, market structure and market access. 

In order to analyse whether the overall back-tested strategy depicts a candidate statistical arbitrage, we implement a test first proposed by \cite{hogan} and further refined by \cite{jarrow}. \cite{hogan} provide a plausible technical definition of statistical arbitrage based on a vanishing probability of loss and variance in the trading profits, and then use this to propose a test for statistical arbitrage using a Bonferroni test (\cite{hogan}). This methodology was extended and generalised by \cite{jarrow} to account for the asymmetry between desirable positive deviations (profits) and undesirable negative deviations (losses), by including a semi-variance hypothesis instead of the originally constructed variance hypothesis, which does not condition on negative incremental deviations. The so-called Min-$t$ statistic is computed, and used in conjunction with a Monte Carlo procedure, to make inferences regarding a carefully defined ``no statistical arbitrage'' null hypothesis. 

This is analogous to evaluating market efficiency in the sense of the Noisy efficient market hypothesis (\cite{black}) whereby a failure to reject the no statistical arbitrage null hypothesis will result in concluding that the market is in fact sufficiently efficient and no persistent anomalies can be consistently exploited by trading strategies over the long term. Traders will always be inclined to employ strategies which depict a statistical arbitrage and especially strategies which have a probability of loss that declines to zero quickly as such traders will often have limited capital and short horizons over which they must provide satisfactory returns (profits) (\cite{jarrow}).

We make the effort here to be very clear that we do not attempt to identify profitable (technical) trading strategies nor to make any claims about the informational value of technical analysis, but rather we will generate a large population of strategies, or experts, constructed from various technical trading rules, and combinations of the associated parameters of these rules, in the attempt to learn something about the aggregate profitability of the population dynamics of the set of experts. 

Expert's will generate trading signals, {\it i.e.} buy, sell or hold decisions for each stock held in their portfolio as based on the underlying parameters and the necessary historic data implied by the parameter's. Once trading signals for the current time period $t$ have been generated by a given expert, a methodology to transform the signals into a set of portfolio weights, or controls, is required. 

We introduce a transformation method that computes controls proportional to the relative volatilities of the stocks for which non-zero trading signals were generated, and then normalise the resulting values such that the self-financing and leverage constraints required by the algorithm are satisfied. The resulting controls are then utilised to compute the corresponding expert wealth's. The experts who accumulate the greatest wealth during a trading period will receive more wealth in the following trading period and thus contribute more to the final aggregated portfolio. This can be best thought of as some sort of ``fund-of-funds'' over the underlying collection of trading strategies. 

This is a meta-expert that aggregates experts that represent all the individual technical trading rules. The overall meta-expert strategy performance is achieved by the online learning algorithm. We explicitly provide equity curves for the individual expert's portfolios, given as the accumulated trading profit through time, along with performance curves for the overall strategy's wealth and the associated profits and losses. 

We perform a back-test of the algorithm on two different data sets over two separate time periods: 1.) one using daily data over a six-year period, and 2.) the other using a mixture of intraday and daily data over a two-month period. A selection of the fifteen most liquid stocks which constitute the Johannesburg Stock Exchange (JSE) Top 40 shares is utilised for the two separate implementations\footnote{More details on the data sets can be found in \cref{sec:data}}.

The overall strategy performance is compared to the BCRP strategy to form a benchmark comparison to evaluate the success of our strategy. The overall strategy is then tested for statistical arbitrage to find that in both a daily, and intraday-daily data implementation, the strategy depicts a statistical arbitrage {\it before costs}. 

A key point here, as in \cite{loonat}, is that it does seem that plausible statistical arbitrages are detected on the meso-scale and in the short-term. However, after accounting for reasonable costs, only the short term trading strategies seem to pass statistical arbitrage tests. This does not imply profitability as these remaining strategies my be structural and hence may not be easily profitably traded out of the system. 

Finally, we analyse the generalisation error of the overall strategy to get a sense of whether or not the strategy conveys back-test overfitting, by estimating the probability of back-test overfitting (PBO) inherent in multiple simulations of the algorithm on subsets of historic data.

The paper will proceed as follows: \cref{sec:OLA} explains the construction of the algorithm including details of how the experts are generated, how their corresponding trading signals are transformed into portfolio weights and a step-by-step break-down of the learning algorithm. In \cref{ssec:statarb}, we introduce the concept of a statistical arbitrage, including the methodology for implementing a statistical arbitrage test, calculating the probability of loss and estimating the PBO for a trading strategy. All experiment results and analyses of implementations of the algorithm are presented in \cref{sec:results}. \cref{sec:conc} states all final conclusions from the experiments and possible future work. 

In summary, we are able to show that on a daily sampled time-scale there is most likely little value in the aggregate trading of the technical strategies of the type considered here. However, on intraday times-scale things look slightly different. Even with reasonable costs accounted for there still seems to be the possibility that price based technical trading cannot be ruled out as offering an avenue for statistical arbitrage. However, considerable care is still required to ensure that one is accounting for the full complexity of market realities that can often make it practically difficult to arbitrage these sorts of apparent trading opportunities out of the market as they may be the results of top-down structure and order-flow itself, rather than some notion of information inefficiency, but are rather the signature of noise trading. 

\section{Learning Technical Trading}\label{sec:OLA}

Rather than using a back-test in the attempt to find the single most profitable strategy, we produce a large population of trading strategies -- ''experts" -- and use an adaptive algorithm to aggregate the performances of the experts to arrive at a final portfolio to be traded. The idea of the online learning algorithm is to consider a population of experts created using a large set of technical trading strategies generated from a variety of parameters and to form an aggregated portfolio of stocks to be traded by considering the wealth performance of the expert population. 

During each trading period, experts trade and execute either a buy (1), sell (-1) or hold (0) actions. These actions are independent of one another and based on each of the individual experts strategies. The trade signals, $\{-1,1,0\}$, are transformed into a set of portfolio weights such that their sum is identical to zero; this ensures that the strategy can be self-funding. We also require that the portfolio is unit leveraged, and hence the absolute sum of controls is equal to one. This is to avoid having to introduce a margin account into the trading mechanics.\footnote{We could in fact have considered leveraged trading but avoided this for simpler trading mechanics. This has no impact on the results of the statistical arbitrage test.}

Based on each individual expert's accumulated wealth up until some time $t$, a final aggregate portfolio for the next period $t+1$ is formed by creating a performance weighted combination of the experts. Experts who perform better in period $t$ will have a larger relative contribution toward the aggregated portfolio to be implemented in period $t+1$ than those who perform poorly. Below, we describe the methodology for generating the expert population. 

\subsection{Expert Generating Algorithm}

\subsubsection{Technical trading}

Technical trading refers to the practice of using trading rules derived from technical analysis indicators to generate trading signals. Here, indicators refer to mathematical formulas based on Open-High-Low-Close (OHLC) price bars, volume traded or a combination of both (OHLCV). An abundance of technical indicators and associated trading rules have been developed over the years with mixed success. Indicators perform differently under different market conditions and different human operators, which is why traders will often use multiple indicators to confirm the signal that one indicator gives on a stock with another indicators signal. Thus, in practice and various studies in the literature, many trading rules generated from indicators are typically back tested on a significant amount (typically thousands of data points) of historical data to find the rules that perform the best\footnote{This is also known as data-mining (\cite{Aronson})}. It is for this reason that we consider a diverse set of technical trading rules. In addition to the set of technical trading strategies, we implement three other popular portfolio selection algorithms each of which has been adapted to generate zero-cost portfolio controls. These three rules generically conform to a combination of chartists trend followers, fundamentalist contrarians, and short-term correlation traders. An explanation of these three algorithms is provided in \cref{ssec:opsrules} while each of the technical strategies are described in \cref{ssec:inds}. 

In order to produce the broad population of experts, we consider combinations among a set of four model parameters. The first of these parameters is the underlying strategy of a given expert, $\boldsymbol{\omega}$, which corresponds to the set of technical trading and trend-following strategies where the total number of different trading rules is denoted by $W$. Each of the rules require at most two parameters each time a buy, sell or hold signal is computed at some time period $t$. The two parameters represent the number of short and long-term look-back periods necessary for the indicators used in the rules. These parameters will determine the amount of historic data considered in the computation of each rule. We will denote the vector of short-term parameters by $\boldsymbol{n_{1}}$ and the long-term parameters by $\boldsymbol{n_{2}}$ which make up two of the four model parameters. Let $L = |\boldsymbol{n_{1}}|$ and $K = |\boldsymbol{n_{2}}|$\footnote{$|\cdot|$ denotes the dimension of a vector} be the number of short-term and long-term look-back parameters respectively. Also, we denote the number of trading rules which utilise one parameter by $W_{1}$ and the number of trading rules utilising two parameters by $W_{2}$ and hence $W = W_{1} + W_{2}$.

The final model parameter, denoted by $\textbf{c}$, refers to object clusters where $\textbf{c}(i)$ is the $i^{th}$ object cluster and $C$ is the number of object clusters. We will consider four object clusters ($C=4$); the trivial cluster which contains all the stocks and the three major sector clusters of stocks on the JSE, namely, Resources, Industrials and Financials.\footnote{See the supplemental material in \cref{ssec:supplemental} for a breakdown of the three sectors into their constituents for daily and intraday data}

The algorithm will loop over all combinations of these four model parameters calling the appropriate strategies, $\boldsymbol{\omega}$, stocks, $\textbf{c}$, and amount of historic data, $\boldsymbol{n_{1}}$ and $\boldsymbol{n_{2}}$, to create a: buy, sell or hold signal at each time period $t$. Each combination of $\boldsymbol{\omega}(i)$ for $i=1, \dots,W$, $\textbf{c}(j)$ for $j=1, \dots, C$, $\boldsymbol{n_{1}}(\ell)$ for $\ell=1, \dots, L$ and $\boldsymbol{n_{2}}(k)$ for $k=1,\dots, K$ will represent an expert. It should be clear that some experts may trade all the stocks, {\it i.e} use the trivial clusters, and others will trade subsets of the stocks, {\it i.e} resources, industrials and financials. It is also important to note that for rules requiring two parameters, the loop over the long-term parameters will only activate at indices $k$ for which $\boldsymbol{n_{1}}(\ell) < \boldsymbol{n_{2}}(k)$ where $\ell$ and $k$ represent the loop index over the short and long-term parameters respectively. The total number of experts, $\Omega$, is then given by
\begin{equation}\nonumber
\begin{split}
\Omega &= \text{no. of experts with 1 parameter} + \text{no. of experts with 2 parameters}\\
&= C\cdot L\cdot W_{1} + C\cdot W_{2} \cdot \sum\Big[\sum(\boldsymbol{n_{2}} > \mbox{max}(\boldsymbol{n_{1}})):\sum(\boldsymbol{n_{2}} > \mbox{min}(\boldsymbol{n_{1}}))]
\end{split}
\end{equation}

We will denote each expert's strategy\footnote{When we refer to strategy, we are talking about the weights of the stocks in the expert's portfolio. As mentioned previously, we will also refer to these weights as controls.} by $\textbf{h}^{n}_{t}$ which is an $(m+1)\times1$ vector representing the portfolio weights of the $n^{th}$ expert for all $m$ stocks and the risk-free asset at time $t$. Here, $m$ refers to the chosen number of stocks to be passed into the expert generating algorithm. As mentioned above, from the set of $m$ stocks, each expert will not necessarily trade all $m$ stocks (unless the expert trades the trivial cluster), since of those $m$ stocks, only a hand full of stocks will fall into a given sector constituency. This implies that even though we specify each expert's strategy ($\textbf{h}^{n}_{t}$) to be an $(m+1)\times1$, we will just set the controls to zero for the stocks which the expert does not trade in their portfolio. Denote the expert control matrix $\textbf{H}_{t}$ made up of all $n$ experts' strategies at time $t$ for all $m$ stocks {\it i.e.} $\textbf{H}_{t} = [\textbf{h}^{1}_{t}, \dots, \textbf{h}^{n}_{t}]$. In order to choose the $m$ stocks to be traded, we take the $m$ most liquid stocks over a specified number of days, denoted by $\delta_{\text{liq}}$. We make the choice of using average daily volume (ADV) as a proxy for liquidity.\footnote{Other indicators of liquidity do exist such as the width of the bid-ask spread and market depth however ADV provides a simple approximation of liquidity.} ADV is simply the average volume traded for a given stock over a period of time. The ADV for stock $m$ over the past $\delta_{\text{liq}}$ periods is
\begin{align}
\mbox{ADV}_{m} =  \frac{1}{\delta_{\text{liq}}}\sum_{t=1}^{\delta_{\text{liq}}}\mbox{V}_{t}^{m}
\vspace{-2mm}
\end{align}
where $\mbox{V}_{t}^{m}$ is the volume of the $m^{th}$ stock at period $t$. The $m$ stocks with the largest ADV will then be fed into the algorithm for trading. 

\subsubsection{Transforming signals into weights}\label{sssec:transform}

In this section, we describe how each individual expert's set of trading signals at each time period $t$ are transformed into a corresponding set of portfolio weights (controls) which constitute the expert's strategy ($\textbf{h}^{n}_{t}$). For the purpose of generality, we refer to the stocks traded by a given expert as $m$ even though the weights of many of these $m$ stocks will be zero for multiple periods as the expert will only be considering a subset of these stocks depending on which object cluster the expert trades. 

Suppose it is currently time period $t$ and the $n^{th}$ expert is trading $m$ stocks. Given that there are $m$ stocks in the portfolio, $m$ trading signals will need to be produced at each trading period. The risk-free assets purpose will be solely to balance the portfolio given the set of trading signals. Given the signals for the current time period $t$ and previous period $t-1$, all hold signals at time $t$ are replaced with the corresponding non-zero signals from time $t-1$ as the expert retains his position in these stocks\footnote{Only the position is retained from the previous period (long/short) not the magnitude of the weight held in the stock}. All non-hold signals at time $t$ are of course not replaced by the previous periods signals as the expert has taken a completely new position in the stock. This implies that when the position in a given stock was short at period $t-1$ for example and the current periods ($t$) signal is long then the  expert takes a long position in the stock rather than neutralising the previous position. Before computing the portfolio controls, we compute a combined signal vector made up of signals from time period $t-1$ and time $t$ using the idea discussed above. We will refer to this combined set of signals as \textit{output signals}. We then consider four possible cases of the output signals at time $t$ for a given expert:

\begin{enumerate}%[topsep=0ex,itemsep=-1ex]
	\item[I.] All output signals are hold (0)
	\item[II.] All output signals are non-negative (0 or 1)
	\item[III.] All output signals are non-positive (0 or -1) 
	\item[IV.] There are combinations of buy, sell and hold signals (0, 1 and -1) in the set of output signals 
\end{enumerate}

Due to the fact that cases II (long-only) and III (short-only) exist, we need to include a risk-free asset in the portfolio so that we can enforce the self-financing constraint; the controls must sum to zero $\sum_{i} w_{i}=0$. We refer to such portfolios as \textit{zero-cost} portfolios. Additionally, we implement a leverage constraint by ensuring that the absolute value of the controls sum to unity: $\sum_{i} \vert w_{i}\vert=1 $.

For case I, we set all stock weights and the risk-free asset weight to zero so that the expert does not allocate any capital in this case since the output signals are all zero. 

For the remaining cases we need to find asset weights that satisfy the investment constraints. From the asset class standard deviations we can define buy (and sell) weight allocation to stocks with positive signals (negative signal) to give by long-asset (short-asset) weights:
\setlength{\abovedisplayskip}{10pt}
\setlength{\belowdisplayskip}{10pt}
\begin{align}\label{eq:sig}
\textbf{w} = \frac{1}{2}\frac{\boldsymbol{\sigma}_{\pm}}{\sum_{i} \boldsymbol{\sigma}_{\pm}(i)}
\end{align}
We will use these weights based on the signals to generate trading positions.

For case II, we compute the standard deviations of the set of stocks which resulted in buy (positive) signals from the output signals using their closing prices over the last 90 days for daily trading and the last 90 trading periods for intraday-daily trading\footnote{See \cref{ssec:int_daily} for details on intraday-daily trading} and use these standard deviations to allocate a weight that is proportional to the volatility (more volatile stocks receive higher weight allocations). 

The number of buy signals, from the set of output signals, can be denoted by $nb$ and the vector of standard deviations of stocks with non-zero output signals is given by $\boldsymbol{\sigma}_{+}$. Then the weight allocated to stocks with positive signals is given by positive signal form of equation (\cref{eq:sig}). Here the lowest value of $\boldsymbol{\sigma}_{+}(i)$ corresponds to the least volatile stock and vice versa for large $\boldsymbol{\sigma}_{+}(i)$. This equation ensures that $\sum_{i}w_{i}= 0.5$. We then short the risk-free asset with a weight of one half $(w_{\mathrm{rf}}=-0.5)$. This allows us to borrow using the risk-free asset and purchase the corresponding stocks within which we take a long position.

Case III is similar to Case II above, however instead of having positive output signals, all output signals are negative. Again, we compute standard deviations of the set of stocks which resulted in sell (negative) signals from the output signals using their closing prices over the last 120 trading periods. Let the number of sell signals from the set of output signals be denoted by $ns$ and denote the vector of standard deviations of stocks to be sold by $\boldsymbol{\sigma}_{-}$. Then the weight allocated to stocks which have short positions is given by the negative signal form of equation (\cref{eq:sig}). We then take a long position in the risk-free asset with a weight of one half $(w_{\mathrm{rf}} = 0.5)$.

For case IV, we use the similar methodology to that discussed above in Case II and III. To compute the weights for the short assets we use the negative signal formula in equation (\cref{eq:sig}); similarly, for the long assets we use positive signal formula from equation (\cref{eq:sig}). We then set the risk-free rate to be equal to $\sum_{i} w_{i}$ in order to enforce the self-financing and fully invested constraints. Finally, assets which had hold signals have their weights are set to zero. 

The method described above is what we will refer to as the \textit{volatility loading} method for transforming signals into controls. A second method is considered, called the \textit{inverse volatility loading} method, and is defined similarly to the method described above, however, instead of multiplying through by the volatility vector in each of the above cases, we multiply through by the inverse of the volatility vector (element-wise inverses). We will not implement the inverse volatility loading method in this study as the results of the two methods are similar. 

%\cref{ssec:olsappend} \Cref{alg:ExpGen} shows the algorithm outline for the Expert Generating Algorithm. The Expert Generating Algorithm calls the \textit{controls} function which transforms trading signals into portfolio controls. The \textit{controls} function is made up of two parts, the first being to compute the output signals as discussed in \Cref{sssec:transform} which is outlined in \cref{ssec:olsappend} \Cref{alg:outputsigs} and the second part is used to transform the output signals into portfolio controls which is outlined in \cref{ssec:olsappend} \Cref{alg:transform}.

\subsection{Online Learning Algorithm}\label{sec:onlinelearn}

Given that we now have a population of experts, each with their own controls, $\textbf{h}^{n}_{t}$, we implement the online learning algorithm to aggregate the expert's strategies at time $t$ based on their performance and form a final single portfolio to be used in the following period $t+1$ which we denote $\textbf{b}_{t}$. The aggregation scheme used is inspired by the Universal Portfolio (UP) strategy taken from the work done by \cite{cover, coverOrd} and a modified version proposed by \cite{gyorfi}. Although, due to the fact that we have several different base experts as defined by the different trading strategies rather than Cover's (see \cite{cover}) constant rebalanced UP strategy, our algorithm is better defined as a meta-learning algorithm (\cite{binli}). We use the subscript $t$ since the portfolio is created using information only available at time $t$ even though the portfolio is implemented in the following time period. The algorithm will run from the initial time $t_{min}$\footnote{This is the time at which trading commences} which is taken to be 2 until terminal time $T$. $t_{min}$ is required to ensure there is sufficient data to compute a return for the first active trading day. We must point out here that experts will only actively begin making trading decisions once there is sufficient data to satisfy their look-back parameter(s) and subsequently, since the shortest look-back parameter is 4 periods, the first trading decisions will only be made during day 5. The idea is to take in $m$ stock's OHLCV values at each time period which we will denote by $\textbf{X}_{t}$. We then compute the price relatives at each time period $t$ given by $\textbf{x}_{t} = (x_{1,t}, \dots, x_{m,t})$ where $x_{m,t}= \frac{P^{c}_{m,t}}{P^{c}_{m,t-1}}$ and where $P^{c}_{m,t}$ is the closing price of stock $m$ at time period $t$. Expert controls are generated from the price relatives for the current period $t$ to form the expert control matrix $\textbf{H}_{t}$. From the corresponding expert control matrix, the algorithm will then compute the expert performance $\textbf{Sh}_{t}$ which is the associated wealth of all $n$ experts at time $t$. Denote the $n^{th}$ expert's wealth at time $t$ by $Sh^{n}_{t}$. We then form the final aggregated portfolio, denoted by $\textbf{b}_{t}$, by aggregating the expert's wealth using the agent mixture update rules.

The relatively simplistic learning algorithm is incrementally implemented online but offline it can be parallelised across experts. Given the expert controls from the Expert Generating Algorithm ($\textbf{H}_{t}$), the online learning algorithm is implemented by carrying out the steps (\cite{loonat}):

\begin{enumerate}
	\item {\bf Update portfolio wealth}:
	Given the portfolio control $b_{m,t-1}$ for the $m^{th}$ asset at time $t-1$, we update the portfolio wealth for the $t^{th}$ period 
	\begin{eqnarray}
	\Delta S_{t} &=& \sum_{m=1}^{M} b_{m,t-1} (x_{m,t+1}-1) +1\\
	S_{t} &=& S_{t-1} \Delta S_{t}
	\end{eqnarray}
	$S_t$ represents the compounded cumulative wealth of the overall aggregate portfolio and $\textbf{S} = S_1,\dots, S_t$ will denote the corresponding vector of aggregate portfolio wealth's over time. Here the realised price relatives for the $t^{th}$ period and the $m^{th}$ asset, $x_{m,t}$, are combined with the portfolio controls for the previous period to obtain the realised portfolio returns for the current period $t$. $\Delta S_{t}-1$ is in fact the profits and losses for the current trading period $t$. Thus, we will use it to update the algorithms overall cumulative profits and losses which is given by 
	\begin{eqnarray}
	\mbox{PL}_{t} &=& \mbox{PL} _{t-1} + \Delta S_{t} - 1 
	\end{eqnarray}
	\item {\bf Update expert wealth}:
	The expert controls $\textbf{H}_{t}$ were determined at the end of time-period $t-1$ for time period $t$ by the expert generating algorithm for $\Omega$ experts and $M$ objects about which the experts make expert capital allocation decisions. At the end of the $t^{th}$ time period the performance of each expert $n$, $Sh^{n}_{t}$, can be computed from the change in the price relatives $x_{m,t}$ for the each of the $M$ objects in the investment universe considered using the closing prices at the start, $P^{c}_{m,t-1}$, and the end of the $t^{th}$ time increment, $P^{c}_{m,t}$, using the expert controls. 
	\begin{eqnarray}
	\Delta Sh^{n}_{t} &=& \left[ {\sum_{m=1}^{M} \textbf{h}^{n}_{t}(x_{m,t}-1)} \right] +1 \\
	Sh^{n}_{t}  &=& Sh^{n}_{t-1} \Delta Sh^{n}_{t}
	\end{eqnarray}
	\item {\bf Update expert mixtures}:
	We consider a UP inspired expert mixture update rule as follows (\cite{cover,gyorfi,loonat}): the mixture of the $n^{th}$ expert for the next time increment, $t+1$, is equivalent to the accumulated expert wealth up until time $t$ and will be used as the update feature for the next unrealised increment subsequent appropriate normalisation
	\be
	q_{n,t+1} = Sh^{n}_{t}
	\ee
	\item {\bf Renormalise expert mixtures}:
	As mentioned previously, we will consider experts such that the leverage is set to unity for zero-cost portfolios: 1.) $\sum_n q_n = 0$ and 2.) $\nu = \sum_n \vert q_n\vert=1$. We will not consider the long-only experts (absolute experts as in \cite{loonat}), but only consider experts whom satisfy the prior two conditions which we will refer to as {\it active} experts. This in fact allows for shorting of one expert against another; then due to the nature of the mixture controls, the resulting portfolio becomes self-funding. 
	\begin{eqnarray}
	q_{n,t+1} = \frac{q_{n,t+1}-\frac{1}{N}\sum_{n=1}^{\Omega} q_{n,t+1}}{\sum_{n=1}^{\Omega} \vert q_{n,t+1}-\frac{1}{N}\sum_{n=1}^{\Omega} q_{n,t+1}\vert}
	\end{eqnarray}
	\item {\bf Update portfolio controls}:
	The portfolio controls $b_{m,t}$ are updated at the end of time period $t$ for time period $t+1$ using the expert mixture controls $q_{n,t+1}$ from the updated learning algorithm and the vector of expert controls $\textbf{h}^{n}_{t}$ for each expert $n$ from the expert generating algorithms using information from time period $t$. We then take a weighted average over all $n$ experts by taking the sum with respect to $n$
	\begin{eqnarray}
	b_{m,t+1} = \sum_n q_{n,t+1} \textbf{h}^{n}_{t}
	\end{eqnarray}  
\end{enumerate}

The strategy is to implement the portfolio controls, wait until the end of the current time increment, measure the features (OHLCV values), update the experts and then re-apply the learning algorithm to compute the expert mixtures and portfolio controls for the next time increment. %More details about the various components of the algorithm are provided in \cref{ssec:olsappend}. Furthermore, a detailed diagram of the MATLAB learning class is illustrated in \cref{fig:algoflow}. 

\subsection{Algorithm implementation for intraday-daily trading}\label{ssec:int_daily}

Intraday trading poses a whole new set of new issues that need to be considered. The first relates to the how the actual trading occurs. The second how to deal with spurious data. It is not as straight-forward as substituting down-sampled transaction data into an algorithm’s built and tested on uniformly sampled daily data. Uniformly sampled daily closing auction data is not equivalent to uniformly sampled intraday bar-data. The end-of-day price discovery process is entirely different to that found intra-day; the prior is a closing auction, the latter has prices being the result of continuous-time trading in a double auction. In addition to this, the first and last data-points of each day lead to over-night gap effects that will lead to spurious signals if not aggregated correctly over multiple days. Rather than dealing with the full complexity of a money management strategy, the main issue that we are concerned with will be the over-night gap effect which relates to the deviation in the prices at the end of day $t-1$ and the start of day $t$. We implement the learning algorithm on a combination of daily and intraday data, whereby decisions made on the daily time scale are made completely independent of those made on the intraday time scale but the dynamics of the associated wealth's generated by the processes are aggregated. We will refer to trading using a combination of daily and intraday data as \textit{intraday-daily} trading. 

The best way to think about it is to consider the experts as trading throughout the day, making decisions based solely on intraday data while compounding their wealth, and once a trading decision is made at the final time bar, the expert makes one last trading decision on that day based on daily historic OHLCV data, where the look-back periods will be based on passed trading days and not on any time bars for that day. The daily trading decision can be thought of as representing the last time bar of the day, where we are just using different data to make the decision. The methodology for each of the intraday and daily trading mechanisms are almost exactly as explained in \cref{sec:onlinelearn} above, however, there are necessary alterations to the algorithm. As in the daily data implementation, a given expert will begin making trading decisions as soon as there is a sufficient amount of data available to them. Here, we begin the algorithm from day two so that there is sufficient data to compute at least one return on the daily time scale. We then loop over the intraday time bars from 9:15am to 4:30pm on each given day. 

To introduce some notation for intraday-daily trading, let $\textbf{Sh}^{F}_{t,t_{I}}$\footnote{Expert wealth is computed as before with $SH^{F}_{n,t,t_{I}}$ = $SH^{F}_{n,t,t_{I}-1}\cdot dSh^{F}_{n,t,t_{I}}$ where $dSh^{F}_{n,t,t_{I}}$ is the $n^{th}$ experts return at time bar $t_{I}$ on day $t$.} be the expert wealth vector for all $n$ experts for the $t_{I}^{th}$ time bar on the $t^{th}$ day and denote by $\textbf{H}^{F}_{t,t_{I}}$ the associated expert control matrix. The superscript $F$ refers to the 'fused' daily and intraday matrices. More specifically, $\textbf{H}^{F}_{t,t_{I}}$ will contain the 88 intraday expert controls followed by the end of day expert controls based on daily closing OHLCV data for each given day over the trading horizon. Denote $T_{I}$ as the final time bar in a day (4:30pm). The $n^{th}$ experts wealth accumulated up until the final time bar $T_{I}$\footnote{$T_{I}$ will always be equal to 88 as there are 88 5-minute time bars between 9:15am and 4:30pm} on day $t$, $Sh^{F}_{n,t,T_{I}+1}$, is calculated from the $n^{th}$ column of the expert control matrix, denoted $\textbf{h}^{F}_{n,t,T_{I}}$, from the previous period $T_{I}$ and is computed solely from intraday data for day $t$. Overall portfolio controls for the final intraday trade on day $t$ ($\textbf{b}_{t,T_{I}+1}$) are computed as before along with the overall portfolio wealth $S_{t,T_{I}+1}$. This position is held until the close of the day's trading when the closing prices of the $m$ stocks $\textbf{P}^{C}_{t}$ are revealed. Once the closing prices are realised, the final intraday position is closed. That is, an offsetting trade of $-\textbf{b}_{t,T_{I}+1}$ is made at the prices $\textbf{P}^{C}_{t}$. This profit/loss is then compounded onto $S_{t,T_{I}+1}$. Thus, no intraday positions are held overnight. The experts will then make one final trading decision based on the daily OHLCV data given that the closing price is revealed and will look-back on daily historic OHLCV data to make these decisions. The $n^{th}$ expert's wealth $Sh^{F}_{n,t,T_{I}+2}$ is updated using controls $\textbf{h}^{F}_{n,t,T_{I}+1}$. The corresponding portfolio controls for all $m$ stocks are computed for the daily trading decision on day $t$ to be implemented at time $t+1$ ($\textbf{b}_{t+1}$), the returns (price relatives) for day $t$ are computed ($\textbf{r}_{t}=\frac{\textbf{P}^{C}_{t}}{\textbf{P}^{C}_{t-1}}$) and the cumulative wealth is $S_{T_{I}+2} = S_{t,T_{I}+1}\cdot(\textbf{b}_{t}\cdot(\textbf{x}_{t}-1) + 1)$ where $S_{t,T_{I}+1} = S_{t,T_{I}}\cdot(\textbf{b}_{T_{I}}(\textbf{x}_{T_{I}}-1))$ with $\textbf{r}_{T_{I}} = \frac{\textbf{P}^{C}_{T_{I}}}{\textbf{P}^{C}_{T_{I}-1}}$. The daily position $\textbf{b}_{t+1}$ is then held until the end of the following day or possibly further into the future (until new daily data portfolio allocations are made). This completes trading for day $t$.

At the beginning of day $t+1$, the expert wealth $Sh^{F}_{n,t+1,1}$\footnote{We do not start the trading day at the first time bar $t_{I}=1$ since we need to compute a return which requires 2 data points.} is set back to unity. Setting experts wealth back to 1 at the beginning of the day, rather than compounding on the wealth from the previous day, is due to the fact that learning on intraday data between days is not possible due to the fact that conditions in the market have completely changed. Trading will begin by computing expert controls $\textbf{h}^{F}_{n,t+1,2}$ for the second time bar, however all experts will not have enough data to begin trading since the shortest look-back parameter is 4 and hence controls will all be set to zero. As the trading day proceeds, experts will begin producing non-zero controls as soon as there is sufficient data to satisfy the amount of data needed for a given look-back parameter. Something to note here is that due to the fact that the STeFI index\footnote{See \cref{sec:data} for more details on the STeFI index} (risk-free asset) is only posted daily, we utilise the same STeFI value for trading throughout the day. Finally, in order to differentiate between daily OHLCV data and intraday OHLCV data, we will denote them as $\textbf{X}_d$ and $\textbf{X}_I$ respectively. %The algorithm outline for intraday-daily trading is illustrated in \cref{ssec:olsappend} \Cref{alg:intra_daily}.

\subsection{Online Portfolio Benchmark Algorithm}\label{ssec:bench}

To get an idea of how well our online algorithm performs, we compare its performance to that of the offline BCRP. As mentioned previously, the hindsight CRP strategy chosen which gives the maximum return of all such strategies in the long run. To find the portfolio controls of such a strategy, we perform a brute force Monte Carlo approach to generate 5000 random CRP strategies on the entire history of price relatives and choose the BCRP strategy to be the one that returns the maximal terminal portfolio wealth. As a note, here, the CRP strategies we consider are long-only. 

\subsection{Transaction Costs and Market Frictions} \label{ssec:transcosts}
Apart from the (direct) transaction fees (commissions) charged by exchanges for the trading of stocks, there are various other costs (indirect) that need to be considered when trading such assets. Each time a stock is bought or sold there are unavoidable costs and it is imperative that a trader takes into account these costs. The other three most important components of these transaction costs, besides commissions charged by exchanges, are the spread\footnote{Spread = best ask price minus best bid price}, price impact and opportunity cost (\cite{spread}).

To estimate indirect transaction costs (TC) for each period $t$, we will consider is the \textit{square-root formula} (\cite{gatheral})
\begin{align}\label{eq:cost}\nonumber
\mbox{TC} = \mbox{Spread Cost} + \sigma.\sqrt{\frac{n}{\mbox{ADV}}} =  M\times \mbox{Spread} + \sigma.\sqrt{\frac{n}{\mbox{ADV}}}
\end{align}
where:
\begin{enumerate}
	\item{\bf Volatility of the returns of a stock ($\sigma$)}: See \cref{ssec:volest} below. 
	\item{\bf Average daily volume of the stock (ADV)}: ADV is computed using the previous 90 days trading volumes for daily trading and the previous 5 days intraday trading volume for intraday-daily trading. 
	\item{\bf Number of shares traded ($n$)}: The number of shares traded ($n$) is taken to be 1bp of ADV for each stock per day for daily trading. The number of stocks traded for intraday-daily trading is assumed to be 70bps of ADV for the entire portfolio per day which is then split evenly among all active trading periods during the day, to get 70bps/85 (assuming 88 5-minute time bars per day) of ADV per stock, per trading period.
	\item{\bf Spread}: Spread is assumed to be 1bps per day (1\%\% /pd) for daily trading. For intraday-daily trading, we assume 20bps per day, which we then split evenly over the day to incur a cost of 0.002/85 per time bar.
	\item{\boldmath{$M$}}: Number of times the trading signal changes from a buy (sell) to a sell (buy,) for all stocks in the portfolio, over consecutive trading periods.\footnote{Every trade with an entry and exit should be assumed to have crossed the spread and as such incurred a direct cost associated with the spread, here estimated as in the above bullet point. The price impact of a trade is assumed to follow the square-root-law, so that the price impact cost of the trade is proportional to the realised volatility across the trade. Here, this is costed as a parent order of size $n$ over each trading period}
\end{enumerate}

The use of the square-root rule in practice dates back many years and is often used as a pre-trade transaction cost estimate (\cite{gatheral}). The first term in \cref{eq:cost} can be regarded as the term representing the \textit{slippage}\footnote{Slippage is often calculated as the difference between the price at which an order for a stock is placed and the price at which the trade is executed}  or \textit{temporary price impact} and results due to our demand for liquidity (\cite{huberman}). This cost will only impact the price at which we execute our trade at and not the market price (and hence the price of subsequent transactions). The second term in \cref{eq:cost} is the (transient) price impact which will not only affect the price of the first transaction but also the price of subsequent transactions by other traders in the market, however, the impact decays over time as a power-law (\cite{Bouchaud}). In the following subsection, we will discuss how the volatility ($\sigma$) is estimated for the square-root formula. Technically, $\sigma$, $n$ and ADV in \cref{eq:cost} should each be defined by a vector representing the volatilities, number of stocks traded and ADV of each stock in the portfolio respectively however for the sake of generality we will write it as a constant thus representing the volatility for a single portfolio stock.

Each trade (say an entry) is consider a child order of the trading strategy across the period. This is not really a faithful representation but was chosen in order to optimise the historic simulation. It should be realised that each child-order is not of a fixed size, as this is determined by the algorithm. However, at the end of the day we have $n$ shares traded, and $M$ entry and exit pairs.

In addition to the indirect costs associated with slippage and price impact as accounted for by the square-root formula, we include direct costs such as the borrowing of trading capital, the cost of regulatory capital and the various fees associated with trading on the JSE (\cite{loonat}). Such costs will also account for small fees incurred in incidences where short-selling has taken place. For the daily data implementation, we assume a total direct cost of 4bps per day. This assumption is purely made to approximately match the total daily transaction cost assumption made by \cite{loonat}. For the intraday-daily implementation a total direct cost of 70bps per day is assumed (following \cite{loonat}) which we then split evenly over each day's active trading periods (85 time bars since first expert only starts trading after the $5^{th}$ time bar) to get a cost of 70bps/85 per period. These costs are indicative and actual price impact requires either real trading experiments or full-scale market simulation - both are intractable in the context of our approach. 

For daily trading, we recover an average daily transaction cost of roughly 17.75bps which is almost double the 10bps assumed by \cite{loonat}. Loonat and Gebbie argue that for intraday trading, it is difficult to avoid a direct and indirect cost of about 50-80bps per day, in each case, leaving a conservative estimate of total costs to be approximately 160bps per day. We realise an overall average cost per period of 2.17bps, while the average cost per day assuming we trade for 85 periods throughout each day is roughly 184bps (85*2.17) for intraday-daily trading.

\subsubsection{Volatility Estimation for Transaction Costs}\label{ssec:volest}
In this section, we will discuss different methods for calculating the estimates for volatility ($\sigma$) for daily and intraday data in the square-root formula (\cref{eq:cost}). 

\paragraph{Daily Data Estimation}

The volatility of daily prices at each day $t$ is taken to be the standard deviation of closing prices over the last 90 days. If 90 days have not passed, then the standard deviation will be taken over the number of days available so far. 

\paragraph{Intraday Data Estimation}

The volatility for each intraday time bar $t_{I}$ on day $t$ is dependent on the time of day. For the first 15 time bars, the volatility is taken to be a forecast of a GARCH(1,1) model which has been fitted on the last 60 returns of the previous day $t-1$. The reason for this choice is that the market is very volatile during the opening hour as well as the fact that there will be relatively few data points to utilise when computing the volatility. The rest of the day's volatility estimates are computed using the Realised Volatility (RV) method (\cite{Andersen}). RV is one of the more popular methods for estimating volatility of high-frequency returns\footnote{Most commonly refers to returns over intervals shorter than one day. This could be minutes, seconds or even milliseconds.} computed from tick data. The measure estimates volatility by summing up intraday squared returns at short intervals (eg. 5 minutes). \cite{Andersen} propose this estimate for volatility at higher frequencies and derive it by showing that RV is an approximate of quadratic variation under the assumption that log returns are a continuous time stochastic process with zero mean and no jumps. The idea is to show that the RV converges to the continuous time volatility (quadratic variation) (\cite{Poon}), which we will now demonstrate. 

Assume that the instantaneous returns of observed log stock prices ($p_{t}$) with unobservant latent volatility ($\sigma_{t}$) scaled continuously through time by a standard Wiener process ($dW_{t}$) can be generated by the continuous time martingale (\cite{Poon})
\begin{align}
dp_{t}=\sigma_{t}dW_{t}.
\end{align}
It follows that the conditional variance of the single period returns, $r_{t+1} = p_{t+1} - p_{t}$ are:
\begin{align}\label{eq:intvol}
\sigma_{t}^{2} = \int_{t}^{t+1} \sigma_{s}^{2}ds.
\end{align}
This is also known as the \textit{integrated volatility} for the period $t$ to $t+1$. Suppose the sampling frequency of the tick data into regularly spaced time intervals is denoted by $f$ so that between period $t-1$ and $t$ there are $f$ continuously compounded returns, then $r_{t+1/f} = p_{t+1/f}-p_{t}$.
%r_{t+2/f} = p_{t+2/f}-p_{t+1/f}
Hence, we can estimate the \textit{Realised Volatility} (RV) based on $f$ intraday returns between periods $t+1$ and $t$ as 
\begin{align}\label{eq:RV}
RV_{t+1} = \sum_{i=1}^{f}r^{2}_{t+i/f}
\end{align}
The argument here is that, provided we sample at frequent enough time steps ($f$), the volatility can be observed theoretically from the sample path of the return process and hence (\cite{karatzas1991brownian,Poon})
\begin{align}
\lim_{f\to\infty} \bigg(\int_{t}^{t+1} \sigma_{s}^{2}ds - \sum_{i=1}^{f}r^{2}_{t+i/f}\bigg) = 0
\end{align}
which says that the RV of a sequence of returns asymptotically approaches the integrated volatility and hence the RV is a reasonable estimate of current volatility levels. 

\section{Testing for Statistical Arbitrage}\label{ssec:statarb}

To test the overall trading strategy for statistical arbitrage, we implement a novel statistical test originally proposed by \cite{hogan} and later modified by \cite{jarrow}, by applying it to the overall strategy's profit and losses $\textbf{PL}$. The idea is to axiomatically define the conditions under which a statistical arbitrage exists and assume a parametric model for incremental trading profits in order to form a null hypothesis derived from the union of several sub-hypotheses which are formulated to facilitate empirical tests of statistical arbitrage. The modified test, proposed by \cite{jarrow}, called the Min-$t$ test, is derived from a set of restrictions imposed on the parameters defined by the statistical arbitrage null hypothesis and is applied to a given trading strategy to test for statistical arbitrage. The Min-$t$ statistic is argued to provide a much more efficient and powerful statistical test compared to the Bonferroni inequality used in \cite{hogan}. The lack of statistical power is reduced when the number of sub-hypotheses increases and as a result, the Bonferroni approach is unable to reject an incorrect null hypothesis leading to a large Type II error. 

To set the scene and introduce the concept of a statistical arbitrage, suppose that in some economy, a stock (portfolio)\footnote{In our study, we will be considering a portfolio} $s_{t}$ and a money market account $B_{t}$\footnote{The money market account is initialised at one unit of a currency {\it i.e.} $B_{0} = 1$.} are traded. Let the stochastic process $(x(t),y(t):t\geq 0)$ represent a zero initial cost trading strategy that trades $x(t)$ units of some portfolio $s_{t}$ and $y(t)$ units of the money market account at a given time $t$. Denote the cumulative trading profits at time $t$ by $V_{t}$. Let the time series of discounted cumulative trading profits generated by the trading strategy be denoted by $\nu(t_{1}), \nu(t_{2}), \dots, \nu(t_{T})$ where $\nu(t_{i}) = \frac{V_{t_{i}}}{B_{t_{i}}}$ for each $i = 1, \dots, T$. Denote the increments of the discounted cumulative profits at each time $i$ by $\Delta \nu_{i} = \nu(t_{i}) - \nu(t_{i-1})$. Then, a statistical arbitrage is defined as:

\begin{definition}[Statistical Arbitrage (\cite{hogan,jarrow})]
	A \textit{statistical arbitrage} is a zero-cost, self-financing trading strategy ($x(t):t\geq 0$) with cumulative discounted trading profits $\nu(t)$ such that:
	\begin{enumerate}
		\item $\nu(0) = 0$,
		\item $\begin{aligned}
		\lim_{t\to\infty} \mathbb{E}^{\mathrm{P}}[\nu(t)] > 0,
		\end{aligned}$
		\item $\begin{aligned}
		\lim_{t\to\infty}   
		\mathrm{P}[\nu(t)<0] = 0, and
		\end{aligned}$
		\item $\begin{aligned}
		\lim_{t\to\infty}   
		Var[\Delta\nu(t) | \Delta\nu(t)<0] = 0.
		\end{aligned}$
	\end{enumerate}\label{def:statarb}
\end{definition}
In other words, a statistical arbitrage is a trading strategy that 1) has zero initial cost, 2) in the limit has positive expected discounted cumulative profits, 3) in the limit has a probability of loss that converges to zero and 4) variance of negative incremental trading profits (losses) converge to zero in the limit. It is clear that deterministic arbitrage stemming from traditional financial
mathematics is in fact a special case of statistical arbitrage (\cite{UCTMphil}).

In order to test for statistical arbitrage, assume that the incremental discounted trading profits evolve over time according to the process
\begin{align}\label{eq:process}
\Delta\nu_{i} = \mu i^{\theta} + \sigma i ^{\lambda}z_{i}
\end{align}
where $i = 1, \dots, T$. There are two cases to consider for the innovations: 1)  $z_{i}$ i.i.d N(0,1) normal uncorrelated random variables satisfying $z_{0}=0$ or 2) $z_{i}$ follows an MA(1) process given by:
\begin{align}
z_{i} = \epsilon_{i} + \phi\epsilon_{i-1}
\end{align}
in which case the innovations are non-normal and correlated. Here, $\epsilon_{i}$ is an i.i.d. N(0,1) normal uncorrelated random variable. It is also assumed that $\Delta \nu_{0} = 0$ and, in the case of our algorithm, $\nu_{t_{min}}$ = 0. We will refer to the first model (normal uncorrelated innovations) as the unconstrained mean (UM) model and the second model (non-normal and correlated innovations) as the unconstrained mean with correlation (UMC) model. Furthermore, we refer to the corresponding models with $\theta=0$ as the constrained mean (CM) and constrained mean with correlation (CMC) respectively, which assume constant incremental profits over time, and hence have an incremental profit process given by: 
\begin{align}\label{eq:CMmodel}
\Delta\nu_{i} = \mu + \sigma i ^{\lambda}z_{i}
\end{align}

The discounted cumulative trading profits for the UM model at terminal time $T$, discounted back to the initial time, which are generated by a trading strategy are given by
\begin{align}\label{eq:disccumprof}
\nu(T) = \sum_{i=1}^{T} \Delta\nu_{i} \sim N\bigg(\mu\sum_{i=1}^{T} i^{\theta} ,\sigma^{2} \sum_{i=1}^{T} i^{2\lambda} \bigg) 
\end{align}
From \cref{eq:disccumprof}, it is straightforward to show that the log-likelihood function for the discounted incremental trading profits is given by:
\begin{align}\label{eq:loglike}
\ell(\mu,\sigma^{2},\lambda,\theta|\Delta\nu) = \ln L(\mu,\sigma^{2},\lambda,\theta\vert\Delta\nu)
= -\frac{1}{2}\sum_{i=1}^{T} \ln(\sigma^{2}i^{2\lambda}) - \frac{1}{2\sigma^{2}} \sum_{i=1}^{T} \frac{1}{i^{2\lambda}} (\Delta\nu_{i} -\mu i^{\theta})^{2}
\end{align}
The probability of a trading strategy generating a loss after $n$ periods is as follows (\cite{jarrow})
\begin{align}\label{eq:probloss}
\mathbf{Pr} \left \{\text{Loss after} \ n \ \text{periods} \right \} = \Phi\Bigg(\frac{-\mu\sum_{i=1}^{n}i^{\theta}}{\sigma(1+\phi)\sqrt{\sum_{i=1}^{n} i^{2\lambda}}}\Bigg)
\end{align}
where $\Phi(\cdot)$ denotes the cumulative standard normal distribution function. For the CM model, \cref{eq:probloss} is easily adjusted by setting $\phi$ and $\theta$ equal to zero. This probability converges to zero at a rate that is faster than exponential. 

As mentioned previously, to facilitate empirical tests of statistical arbitrage under \Cref{def:statarb}, a set of sub-hypotheses are formulated to impose a set of restrictions on the parameters of the underlying process driving discounted cumulative incremental trading profits and are as follows:

\begin{proposition}[UM Model Hypothesis (\cite{jarrow})]\label{thm:hypoth}
	Under the four axioms defined in \Cref{def:statarb}, a trading strategy generates a statistical arbitrage under the UM model if the discounted incremental trading profits satisfy the intersection of the following four sub-hypotheses jointly: i.) $H_{1}: \mu > 0$, ii.) $H_{2}: -\lambda > 0$ or $\theta-\lambda > 0$, iii.)$H_{3}: \theta-\lambda+\frac{1}{2} > 0$, and  $H_{4}: \theta + 1 >0$.
\end{proposition}

An intersection of the above sub-hypotheses defines a statistical arbitrage, and as by De Morgan's Laws\footnote{This states that the complement of the intersection of sets is the same as the union of their complements.}, the null hypothesis of no statistical arbitrage is defined by a union of the sub-hypotheses. Hence, the no statistical arbitrage null hypothesis is the set of sub-hypotheses which are taken to be the complement of each of the sub-hypotheses in \Cref{thm:hypoth}:

\begin{proposition}[UM Model Alternative Hypothesis (\cite{hogan,jarrow})]
	Under the four axioms defined in \Cref{def:statarb}, a trading strategy does not generate a statistical arbitrage if the discounted incremental trading profits satisfy any one of the following four sub-hypotheses: i.) $H_{1}: \mu \leq 0$, ii.) $H_{2}: -\lambda \leq 0$ or $\theta-\lambda \leq 0$, iii.) $H_{3}: \theta-\lambda+\frac{1}{2} \leq 0$ , and iv.) $H_{4}: \theta + 1 \leq 0$ 
\end{proposition}\label{prop:hypothC}

The null hypothesis is not rejected provided that a single sub-hypothesis holds. The Min-$t$ test is then used to test the above null hypothesis of no statistical arbitrage by considering each sub-hypothesis separately using the t-statistics $t(\hat{\mu}), t(-\hat{\lambda}), t(\hat{\theta}- \hat{\lambda}), t(\hat{\theta} - \hat{\lambda} + 0.5),$ and $t(\hat{\theta} + 1)$, where the hats denote the Maximum Likelihood Estimates (MLE) of the parameters. The Min-$t$ statistic is defined as (\cite{jarrow})
\begin{align}\label{eq:Min-t}
\text{Min-$t$} = \text{Min}\{&t(\hat{\mu}), t(\hat{\theta}- \hat{\lambda}), t(\hat{\theta} - \hat{\lambda} + 0.5), \text{Max}[t(-\hat{\lambda}), t(\hat{\theta} + 1)]\}
\end{align}
The intuition is that the Min-$t$ statistic returns the smallest test statistic which is the sub-hypothesis which is closest to being accepted. The no statistical arbitrage null is then rejected if Min-$t$ $>$ $t_{c}$ where $t_{c}$ depends on the significance level of the test which we will refer to as $\alpha$. Since the probability of rejecting cannot exceed the significance level $\alpha$, we have the following condition for the probability of rejecting the null at the $\alpha$ significance level
\begin{align}\label{eq:probreject}
\mathbf{Pr}\{\text{Min-}t > t_{c} \vert \mu, \lambda, \theta, \sigma\} \leq \alpha
\end{align}
What remains is for us to compute the critical value $t_{c}$. We will implement a Monte Carlo simulation procedure to compute $t_{c}$ which we describe in more detail in \cref{ssec:statarboutline} step \ref{tc} below. 

\subsection{Outline of the Statistical Arbitrage Test Procedure}\label{ssec:statarboutline}

The steps involved in testing for statistical arbitrage are outlined below:
\begin{enumerate}
	\item{\bf Trading increments $\Delta\nu_{i}$}:
	From the vector of cumulative trading profits and losses, compute the increments $(\Delta\nu_{1}, \dots, \Delta\nu_{T})$ where $\Delta\nu_{i} = \nu(t_{i}) - \nu(t_{i-1})$. 
	\item{\bf Perform MLE}:
	Compute the likelihood function, as given in \cref{eq:loglike}, and maximise it to find the estimates of the four parameters, namely, $\hat{\mu}, \hat{\sigma}, \hat{\theta}$ and $\hat{\lambda}$. The log-likelihood function will obviously be adjusted depending on whether the CM $(\theta=0)$ or UM test is implemented. We will only consider the CM test in this study. Since MATLAB's built-in constrained optimization algorithm\footnote{Here, we are referring to MATLAB's \textit{fmincon} function} only performs minimization, we minimise the negative of the log-likelihood function {\it i.e.} maximise the log-likelihood. 
	\item{\bf Standard errors}:
	From the estimated parameters in the MLE step above, compute the negative Hessian estimated at the MLE estimates which is indeed the Fisher Information (FI) matrix denoted by $\textbf{I}(\Theta)$. In order to compute the Hessian, the analytical partial derivatives are derived from \cref{eq:loglike}. Standard errors are then taken to be the square roots of the diagonal elements of the inverse of $\textbf{I}(\Theta)$ since the inverse of the Fisher information matrix is an asymptotic estimator of the covariance matrix. 
	\item{\bf Min-$t$ statistic}:
	Compute the t-statistics for each of the sub-hypotheses which are given by $t(\hat{\mu}), t(-\hat{\lambda}), t(\hat{\theta}- \hat{\lambda}), t(\hat{\theta} - \hat{\lambda} + 0.5),$ and $t(\hat{\theta} + 1)$ and hence the resulting Min-$t$ statistic given by \cref{eq:Min-t}. Obviously, $t(\hat{\theta}- \hat{\lambda})$, $t(\hat{\theta} - \hat{\lambda} + 0.5)$ and $t(\hat{\theta} + 1)$ will not need to be considered for the CM test. 
	\item{\bf Critical values}:
	Compute the critical value at the $\alpha$ significance level using the Monte Carlo procedure (uncorrelated normal errors) and Bootstrapping (correlated non-normal errors) \label{tc} 
	\begin{enumerate}[label = (\alph*), ref=\theenumi{} (\alph*)]
		\item{\bf CM model}\label{CMtc} 
		First, simulate 5000 different profit process using \cref{eq:CMmodel} with $(\mu,\lambda,\sigma^{2}) = (0,0,0.01)$\footnote{$t_{c}$ is maximised when $\mu$ and $\lambda$ are zero. $\sigma^{2}$ is set equal to 0.01 to approximate the empirical MLE estimate (\cite{jarrow}).}. For each of the 5000 profit processes, perform MLE to get estimated parameters, the associated t-statistics and finally the Min-$t$ statistics. $t_{c}$ is the taken to be the 1-$\alpha$ quantile of the resulting distribution of Min-$t$ values. 
	\end{enumerate}
	\item{\bf P-values}:	
	Compute the empirical probability of rejecting the null hypothesis at the $\alpha$ significance level using \cref{eq:probreject} by utilising the critical value from the previous step and the simulated Min-$t$ statistics. 
	\item{\bf n-Period Probability of Loss}: \label{ploss}
	Compute the probability of loss after $n$ periods for each $n = 1, \dots, T$ and observe the number of trading periods it takes for the probability of loss to converge to zero (or below 5\% as in the literature). This is done by computing the MLE estimates for the vector ($\Delta\nu_{1}, \Delta\nu_{2}, \dots \Delta\nu_{n}$) for each given $n$ and substituting these estimates into \cref{eq:probloss}. 
\end{enumerate}

\subsection{Estimates of the Probability of Back-test Overfitting (PBO)}
\label{ssec:pbo}

%When measuring the performance of a back-tested strategy, there are two 
%measurements: in-sample (IS) performance and out-of-sample (OOS) performance. 
%IS performance is simulated over a sample of data used in the design of the 
%trading strategy which can be referred to as the “training set". OOS 
%performance is simulated of the sample of data used to test the trading 
%strategy which is also known as the “testing set". 
\cite{bailey} heavily criticise recent studies which claim 
to have designed profitable investment or trading strategies since many of these studies are only based on in-sample (IS) statistics, without evaluating out-of-sample (OOS) performance. We briefly addressed this concern by computing estimate of the probability of back-test overfitting (PBO) using the combinatorially symmetric cross-validation (CSCV) procedure outlined in \cite{pbo}. Typically, an investor/researcher will run many ($N$) trial back-tests to select the parameter combinations which optimise the performance of the algorithm (usually based on some performance evaluation criterion such as the Sharpe Ratio). The idea is to perform CSCV on the matrix of performance series over time of length $T_{\text{BL}}$\footnote{The subscript BL stands for back-test length} for the $N$ separate trial simulations of the algorithm. 

Here, we must be clear that when we refer to IS, we do not mean the ``training set'' per say, during which the moving average look-back parameters were calculated for example. Rather, we refer to IS as being the subset of observations utilised in selecting the optimal strategy from the $N$ back-test trials.

In the case of the algorithm proposed in this study, since the large set of trialled parameters form the basis of the learning algorithm in the form of the experts, we cannot observe the effect of different parameters settings on the overall strategy, as these are already built into the underlying algorithm. Rather, we will run $N$ trial back-test simulations on independent subsets of historical data to get an idea of how the algorithm performs on different subsets of unseen data. We can then implement the CSCV procedure on the matrix of profits and losses resulting from the trials to recover a PBO estimate. Essentially, there is no training of parameters taking place in our model, as all parameter combinations are considered, and the weights of the performance weighted average of the expert's strategies associated with the different parameter combinations are ``learnt''.

More specifically, we choose a back-test length $T_{\text{BL}}$ for each subset and split the entire history of OHLCV data into subsets of this length. The learning algorithm is then implemented on each subset to produce $N=\lfloor T/T_{\text{BL}} \rfloor$ profit and loss time series. Note that the subsets will be completely independent from one another as there is no overlapping of the data that each separate simulation is run on. The results from the simulations are presented in \cref{tab:pbo} below.

\LTcapwidth=9cm
\begin{longtable}{p{2.5cm}p{1cm}p{1.8cm}p{1.2cm}}
		\caption{Number of back-test trials ($N$), back-test length for each 
		simulation ($T_{\text{BL}}$) and the resulting PBO estimates for the daily 
		and intraday-daily implementation.}\label{tab:pbo}\\
	\hline
		\rule{0pt}{2.5ex}
	&  $N$    &      $T_{BL}$ &  PBO   \\
	\hline
	\rule{0pt}{3ex}	
	\noindent Daily & 30     & 60 days &        1.4\% \\ 
	Intraday-daily & 22    & 3 days &      11.4\%  \\
	\hline
\end{longtable}

\section{The Data}\label{sec:data}

\subsection{Daily Data}
The daily data is sourced from Thomson Reuters and contains data corresponding to all stocks listed on the JSE Top 40\footnote{Please refer to the supplemental material in \cref{ssec:supplemental} for the full names of the Bloomberg ticker symbols.}. The data set consists of data for 42 stocks over the period 01-01-2005 to 29-04-2016 however we will only utilise the stocks which traded more than 60\% of the time over this period. Removing such stocks leaves us with a total of 31 stocks. The data comprises of the opening price $(P^{o})$, closing price $(P^{c})$, lowest price $(P^{l})$, the highest price $(P^{h})$ and daily traded volume $(V)$ (OHLCV). In additions to these 31 stocks, we also require a risk-free asset for balancing the portfolio. We make the choice of trading the Short Term Fixed Interest (STeFI) index. The STeFI benchmark is a proprietary index that measures the performance of Short Term Fixed Interest or money market investment instruments in South Africa. It is constructed by Alexander Forbes (and formerly by the South African Futures Exchange (SAFEX)) and has become the industry benchmark for short-term cash equivalent investments (up to 12 months) (\cite{stefi}).

\subsection{Intraday-Daily Data}
Bloomberg is the source of all tick (intraday) data used in this paper. The data set consists of 30 of the Top 40 stocks on the JSE from 02-01-2018 to 29-06-2018. The data is then sampled at 5-minute intervals to create an OHLCV entry for all 5-minute intervals over the 6-month period. We remove the first 10 minutes and last 20 minutes of the continuous trading session (9:00-16:50) as the market is relatively illiquid and volatile during these times which may lead to spurious trade decisions. We are thus left with 88 OHLCV entries for each stock on any given day. In addition to the intraday data, daily OHLCV data for the specified period is required for the last transaction on any given day. As in the daily data case, we make use of the STeFI index as the risk-free asset, and hence the daily entries for the STeFI index are included in this data set. The data was sourced from a Bloomberg terminal using the R Bloomberg API, \textit{Rblpapi}, and all data processing is done in MATLAB to get the data into the required form for the learning algorithm. 

\section{Results and Analysis}\label{sec:results}
\subsection{Daily Data}
In this section, we implement the various algorithms described above in order to plot a series of graphs for daily JSE Top 40 data as discussed in \cref{sec:data} above. We will plot five different graphs: first is the overall portfolio wealth over time which corresponds to $S_{t}$ as described above, second, the cumulative profit and losses over time $\mbox{PL}_{t}$, third, the relative population wealth of experts corresponds to the wealth accumulated over time by each of the experts competing for wealth in the algorithm $\textbf{Sh}_{t}$ and finally, the relative population wealth of the strategies which takes the mean over all experts for each given trading strategy to get an accumulated wealth path for each technical trading rule.

For the purpose of testing the learning algorithm, we will identify the 15 most liquid stocks over one year prior to the start of active trading. The stocks, ranked by liquidity, are as follows: FSRJ.J, OMLJ.J, CFRJ.J, MTNJ.J, SLMJ.J, NTCJ.J, BILJ.J,	SBKJ.J,	WHLJ.J,	AGLJ.J,	SOLJ.J,	GRTJ.J,	INPJ.J,	MNDJ.J and RMHJ.J.

\subsubsection{No Transaction Costs}
\begin{figure}
	\centering
	\includegraphics[width=7cm]{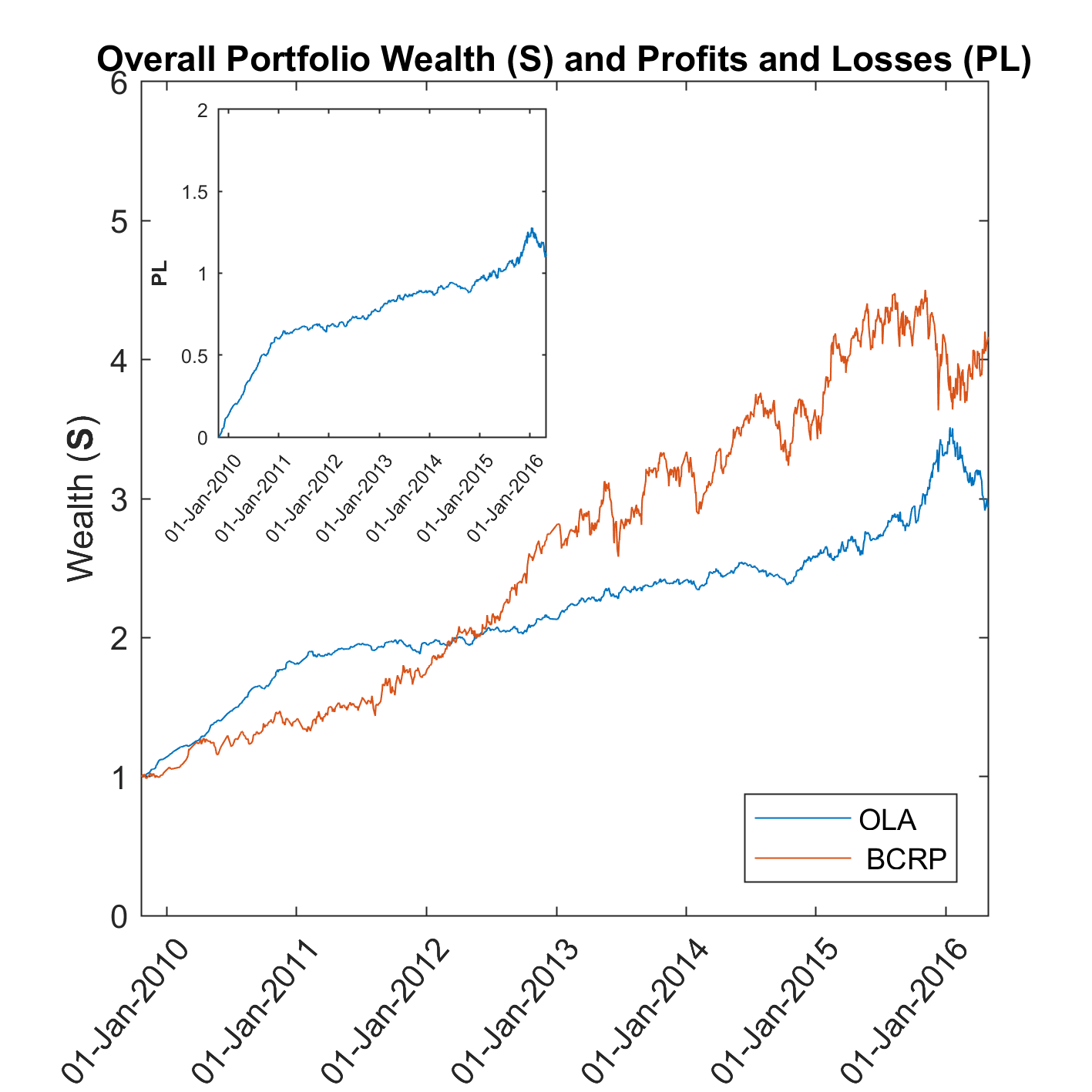}
	\caption{Overall cumulative portfolio wealth ($S$) for daily data with no transaction costs (blue) and the benchmark BCRP strategy (orange). The figure inset illustrate the associated profits and losses (\textbf{PL}) of the strategy.}\label{fig:20_strats16_22-Oct-2009-29-Apr-2016_3001_S}
\end{figure}
\begin{figure}
	\centering
	\subfigure[]{\includegraphics[width=7.5cm]{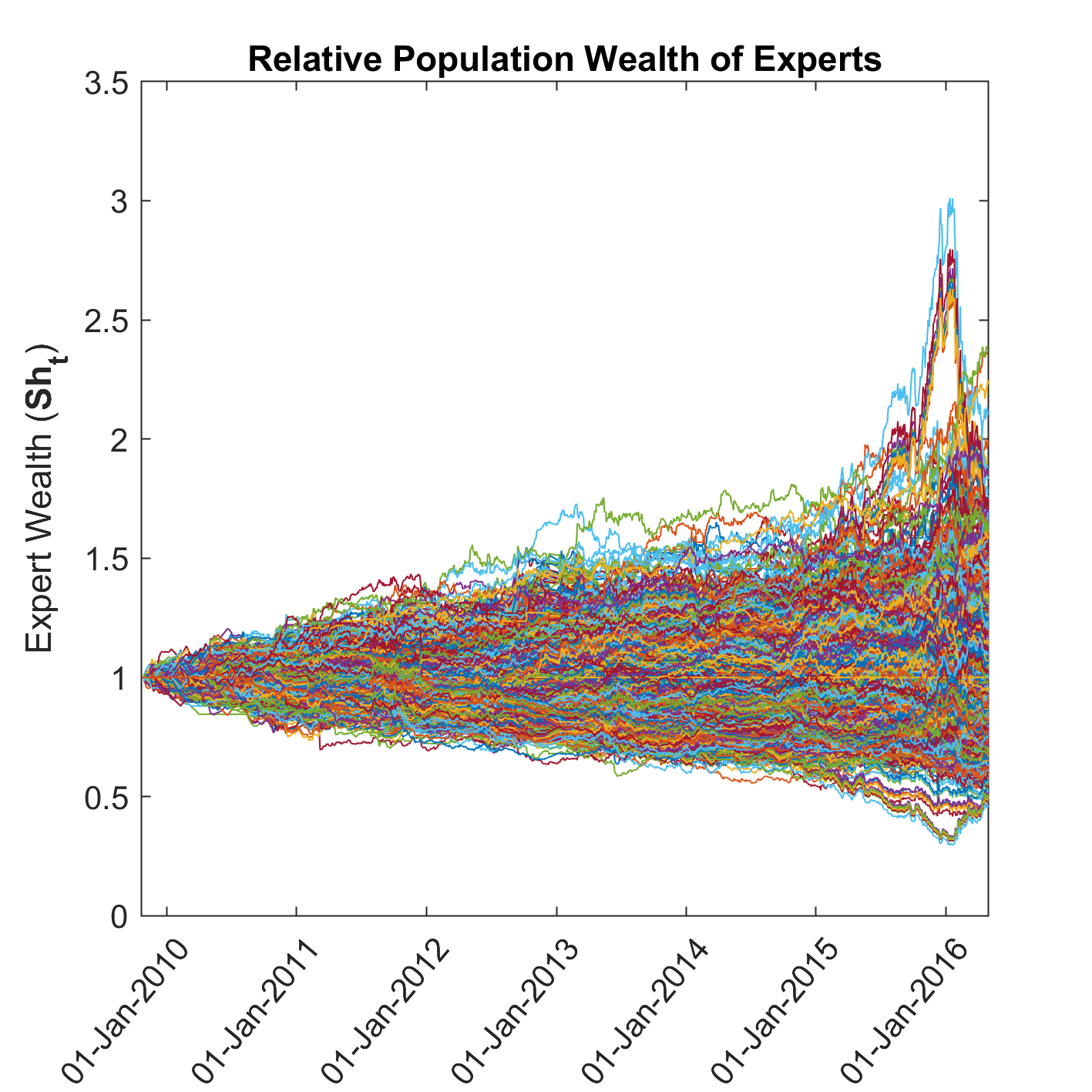}\label{fig:20_strats16_10-Nov-2010-29-Apr-2016_3001_SH}}
	\quad
	\subfigure[]{\includegraphics[width=7.5cm]{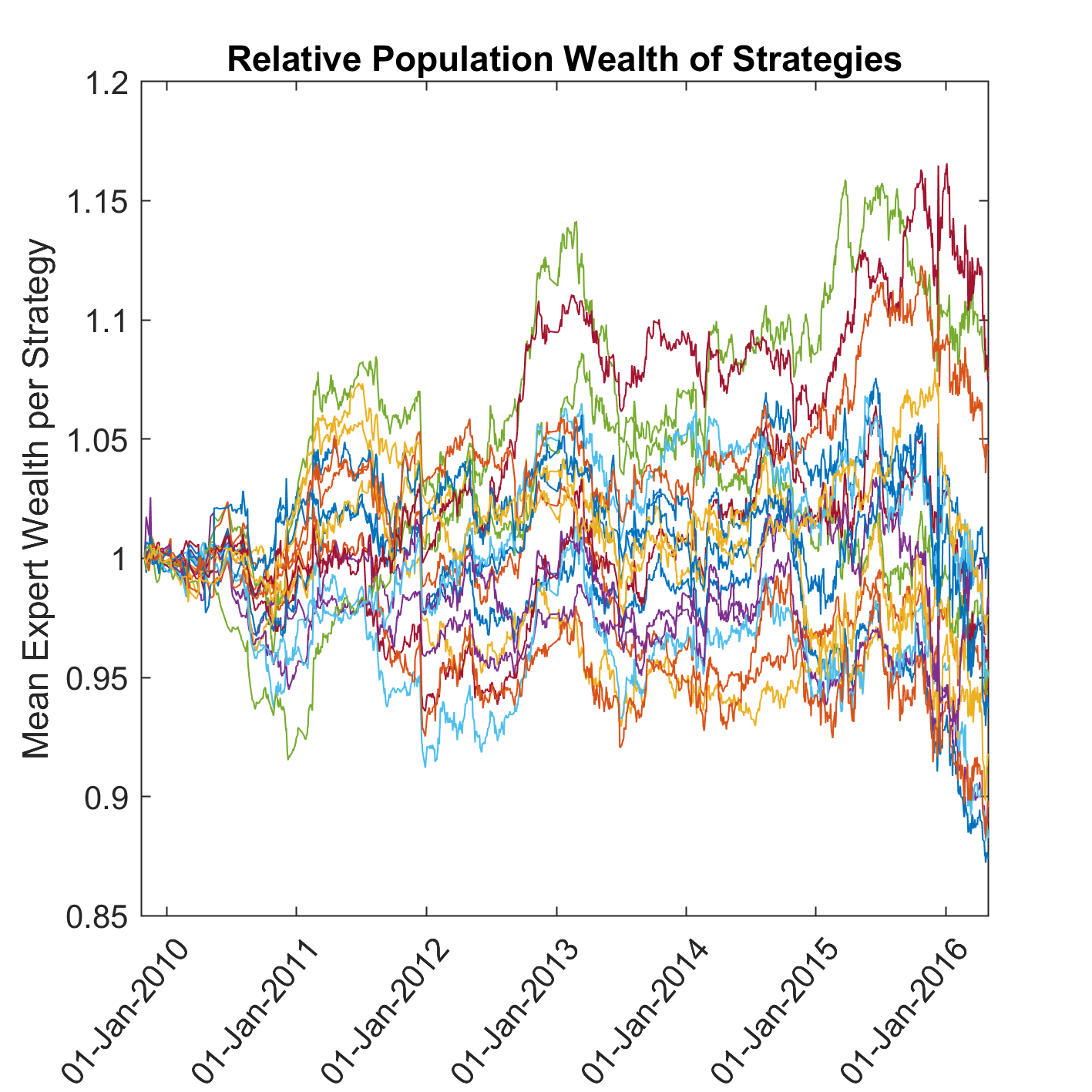}\label{fig:20_strats16_10-Nov-2010-29-Apr-2016_3001_SHave}}
	\caption{\cref{fig:20_strats16_10-Nov-2010-29-Apr-2016_3001_SH} illustrates the expert wealth ($\textbf{Sh}$) for all $\Omega$ experts for daily data with no transaction costs. \cref{fig:20_strats16_10-Nov-2010-29-Apr-2016_3001_SHave} illustrates the mean expert wealth of all experts for each trading strategy ($\boldsymbol{\omega}(i)$) for daily data with no transaction costs.}
\end{figure}

Barring transaction costs, it's clear that the portfolio makes favourable cumulative returns on equity over the six-year period as is evident in \cref{fig:20_strats16_22-Oct-2009-29-Apr-2016_3001_S}. The performance of the online learning algorithm (blue) is similar to that of the benchmark BCRP strategy (orange) which is promising as the original literature proves that the algorithm should track such a benchmark in the long-run. The figure inset in \cref{fig:20_strats16_22-Oct-2009-29-Apr-2016_3001_S} illustrates that the overall strategy provides consistent positive trading profits over the entire trading horizon. \cref{fig:20_strats16_10-Nov-2010-29-Apr-2016_3001_SH} shows the expert wealth for all $\Omega$ experts and \cref{fig:20_strats16_10-Nov-2010-29-Apr-2016_3001_SHave} shows the mean expert wealth for each strategy. These figures show that on average, the underlying experts perform fairly poorly compared to the overall strategy however there is evidence that some experts make satisfactory returns over the period.

\cref{tab:stats_wealth} and \cref{tab:stats_pnl} provide the group summary statistics of the terminal wealth's of experts and of the expert's profits and losses over the entire trading horizon respectively where experts are grouped based on their underlying strategy $\boldsymbol{\omega}(i)$. The online Z-Anticor\footnote{Please refer to \cref{ssec:inds} and \cref{ssec:opsrules} for a detailed description of the various trading rules mentioned in the \cref{tab:stats_wealth} and \cref{tab:stats_pnl}.} algorithm produces the best expert (maximum terminal wealth) followed closely by the slow stochastic rule while Z-Anticor also produces experts with the greatest mean terminal wealth over all experts (column 2). Additionally, Z-Anticor produces expert's with wealth's that vary the most (highest standard deviation). Williams \%R produces the worst expert by quite a long way (minimum terminal wealth). The trading rule with the lowest mean terminal wealth and worst mean ranking are SAR and slow stochastic respectively. With regards to the expert's profits and losses (\cref{tab:stats_pnl}), the momentum rule (MOM) produces the expert with the greatest profit in a single period. SAR followed by Anti-Z-BCRP produce the worst and second worst mean profit/loss per trading period respectively whereas Z-Anticor and Z-BCRP achieve the best mean profit/loss per trading period. 
\LTcapwidth=14cm
\begin{longtable}{p{4cm}p{3.2cm}p{2cm}p{1.5cm}p{1cm}}
		\caption{Group summary statistics of the overall rankings of experts grouped by their underlying strategy ($\boldsymbol{\omega}(i)$ where $i= 1, \dots, 17$) for the daily trading. In brackets next to mean are the mean overall ranking of experts within the group of their underlying strategy.}\label{tab:stats_wealth}\\
		\hline
		\rule{0pt}{3ex}
		Strategy & Mean (mean rank) & St. Dev. & Min  & Max  \\ %Best expert wealth & Worst expert wealth \\
		\hline
		\rule{0pt}{3ex}
		\noindent EMA X-over& 0.8739 (673.6343)	 &	0.1767		&	0.5216		& 1.4493\\
		Ichimoku Kijun Sen&  0.9508 (623.3194) & 0.2313&		0.5424&		1.5427\\
		MACD &	0.9504 (657.7639) &	0.1750&	0.5601&	1.6065\\
		Moving Ave X-over & 0.8895 (632.6944) &	0.1930&	0.5206&	1.4505\\
		ACC & 1.0994 (736.5833) &	0.3131&	0.5283&	1.9921\\
		BOLL & 1.0499 (569.1944) &	0.3536&	0.6076&	1.7746\\
		Fast Stochastic& 0.9995 (778.6111) &	0.3699&	0.6006&	1.8555\\
		MARSI & 1.0723 (639.3611) &	0.2081&	0.6947&	1.6917\\
		MOM & 1.0403 (681.4444) &	0.1353&	0.7349&	1.3595\\
		Online Anti-Z-BCRP& 0.7579 (731.9444)&	0.1935&	0.4649&	1.0924\\
		Online Z-Anticor & 1.3155 (694.5278) &	0.4388&	0.6363&	2.3886\\
		Online Z-BCRP &	1.2818 (652.8611)&	0.2637&	0.8561&	1.8341\\
		PROC & 	0.8963 (718.0833) &	0.1631&	0.6305&	1.2161\\
		RSI & 1.1339 (757.3889) &	0.2544&	0.6440&	1.7059\\
		SAR & 0.7314 (654.1111) &	0.0619&	0.6683&	0.8683\\
		Slow Stochastic & 1.1135 (793.2222) &	0.3302&	0.6955&	2.1023\\
		Williams \%R & 	0.9416 (728.6944) &	0.3150&	0.4662&	1.5131\\
		\hline
\end{longtable}

\begin{longtable}[t]{p{4cm}p{3.2cm}p{2cm}p{1.5cm}p{1cm}}
      \caption{Group summary statistics of the expert's profits and losses per period grouped by their underlying strategy ($\boldsymbol{\omega}(i)$ where $i= 1, \dots, 17$).}\label{tab:stats_pnl}\\
		\hline
		\rule{0pt}{3ex}
		Strategy & Mean & St. Dev. & Min  & Max   \\
		\hline
		\rule{0pt}{3ex}
		\noindent EMA X-over&	-0.00010&	0.00633&	-0.09745&	0.08074\\
		Ichimoku Kijun Sen&	-0.00004&	0.00723&	-0.10467&	0.06157\\
		MACD&	-0.00003&	0.00725	&-0.15993&	0.08074\\
		Moving Ave X-over&	-0.00009&	0.00644	&-0.15993&	0.11482\\
		ACC&	0.00007	&0.00760&	-0.15993&	0.08028\\
		BOLL&	0.00002	&0.00711	&-0.06457&	0.06480\\
		Fast Stochastic	&-0.00001	&0.00847&	-0.06469&	0.06279\\
		MARSI&	0.00006	&0.00612&	-0.06788&	0.06527\\
		MOM	&0.00004&	0.00603&	-0.06051&	0.15820\\
		Online Anti-Z-BCRP&	-0.00022&	0.00773&	-0.09847&	0.09336\\
		Online Z-Anticor&	0.00021&	0.00759	&-0.06475&	0.09773\\
		Online Z-BCRP&	0.00021	&0.00771&	-0.09336&	0.09847\\
		PROC&	-0.00007&	0.00733	&-0.10467	&0.09745\\
		RSI	&0.00010&	0.00666&	-0.06460&	0.09745\\
		SAR	&-0.00023&	0.00724&	-0.10467&	0.08724\\
		Slow Stochastic&	0.00009	&0.00809&	-0.06480&	0.06820\\
		Williams \%R&	-0.00006&	0.00815	&-0.06820&	0.06317\\
		\hline
	\end{longtable}

\begin{figure}
	\centering
	\subfigure[]{\includegraphics[width=7cm]{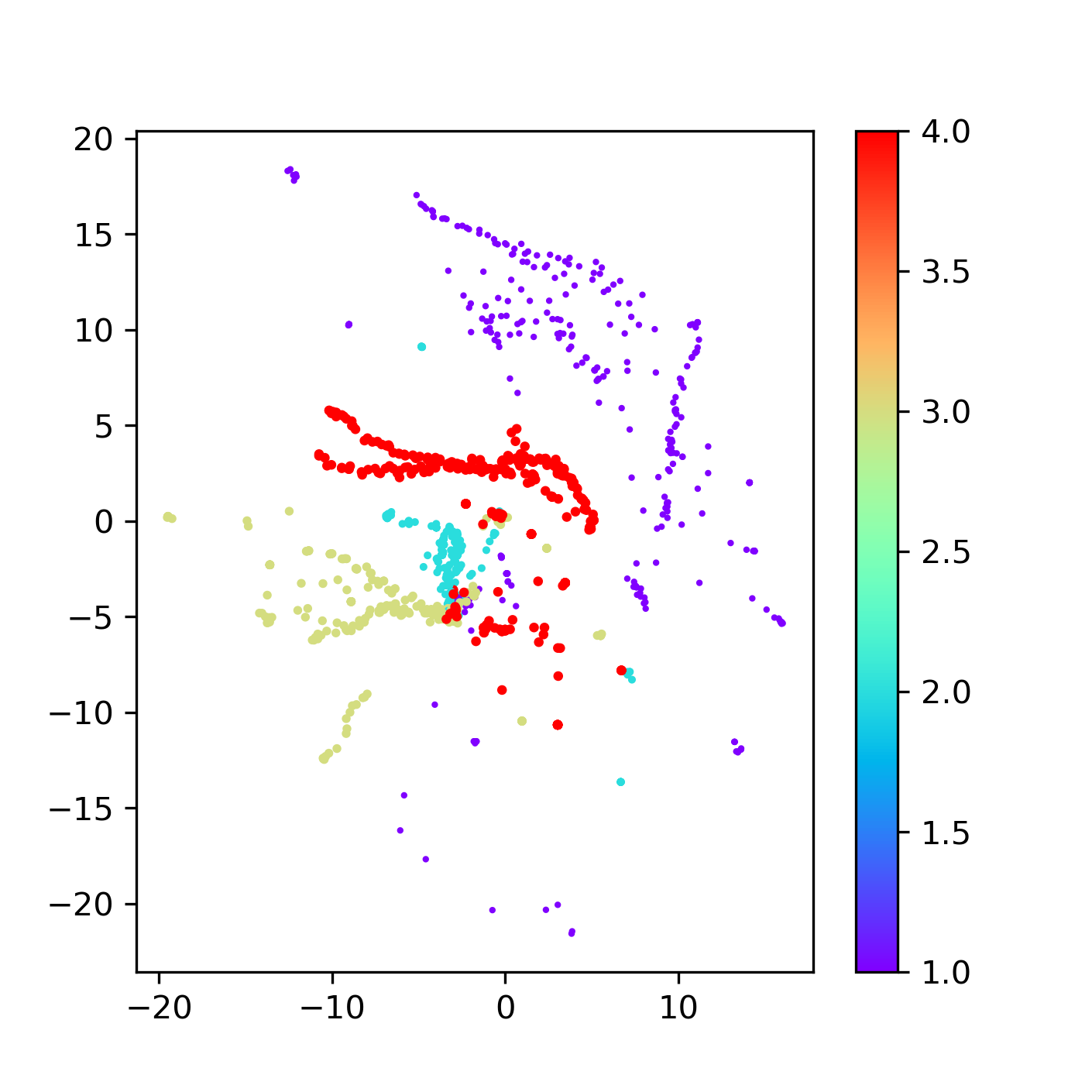}\label{fig:VAE}}\quad
	\subfigure[]{\includegraphics[width=7cm]{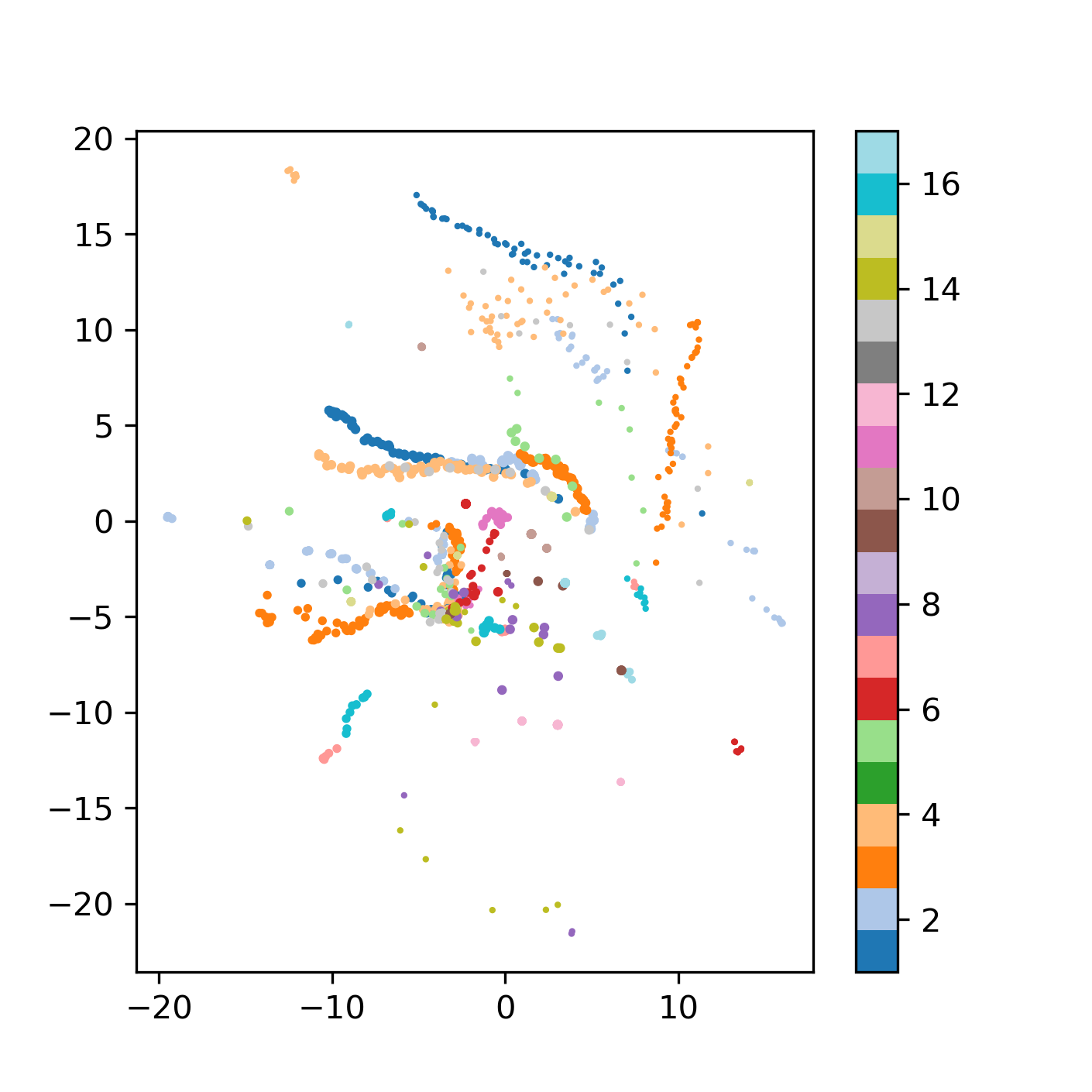}\label{fig:VAEstrat}}
	\caption{\cref{fig:VAE} and \cref{fig:VAEstrat} show the latent space of Variational Autoencoder on the time series' of expert wealth's implemented using Keras in Python. In \cref{fig:VAE} experts are coloured by which of the 4 object clusters they trade whereas in \cref{fig:VAEstrat}, experts are coloured by their underlying trading strategy $\boldsymbol{\omega}(i)$.}
\end{figure}

\cref{fig:VAE} illustrates the 2-D plot of the latent space of a Variational Autoencoder (VAE) for the time series' of wealth's of all the experts with experts coloured by object cluster. It is not surprising that the expert's wealth time series' show quite well-defined clusters in terms of the stock which experts trade in their portfolio as the stocks that each expert trades will be directly related to the decisions they make given the incoming data and hence the corresponding returns (wealth) they achieve. 

To provide some sort of comparison, in \cref{fig:VAEstrat} we plot the same results as above but this time we colour the experts in terms of their underlying strategy $\boldsymbol{\omega}(i)$. The VAE seems to be able to pick up much clearer similarities (dissimilarities) between the experts based on the stocks they trade compared to which strategy they utilise providing evidence that the achieved wealth has a much stronger dependence on the stock choice rather than the chosen strategy. This may be an important point to consider and gives an indication that it may be worth considering more sophisticated ways to choose the stocks to trade rather than developing more sophisticated/profitable strategies. A discussion on the features that should be considered by a quantitative investment manager in assessing an assets usefulness is provided in \cite{aseetuseful}.  

\begin{figure}
	\centering
	\includegraphics[width=7cm]{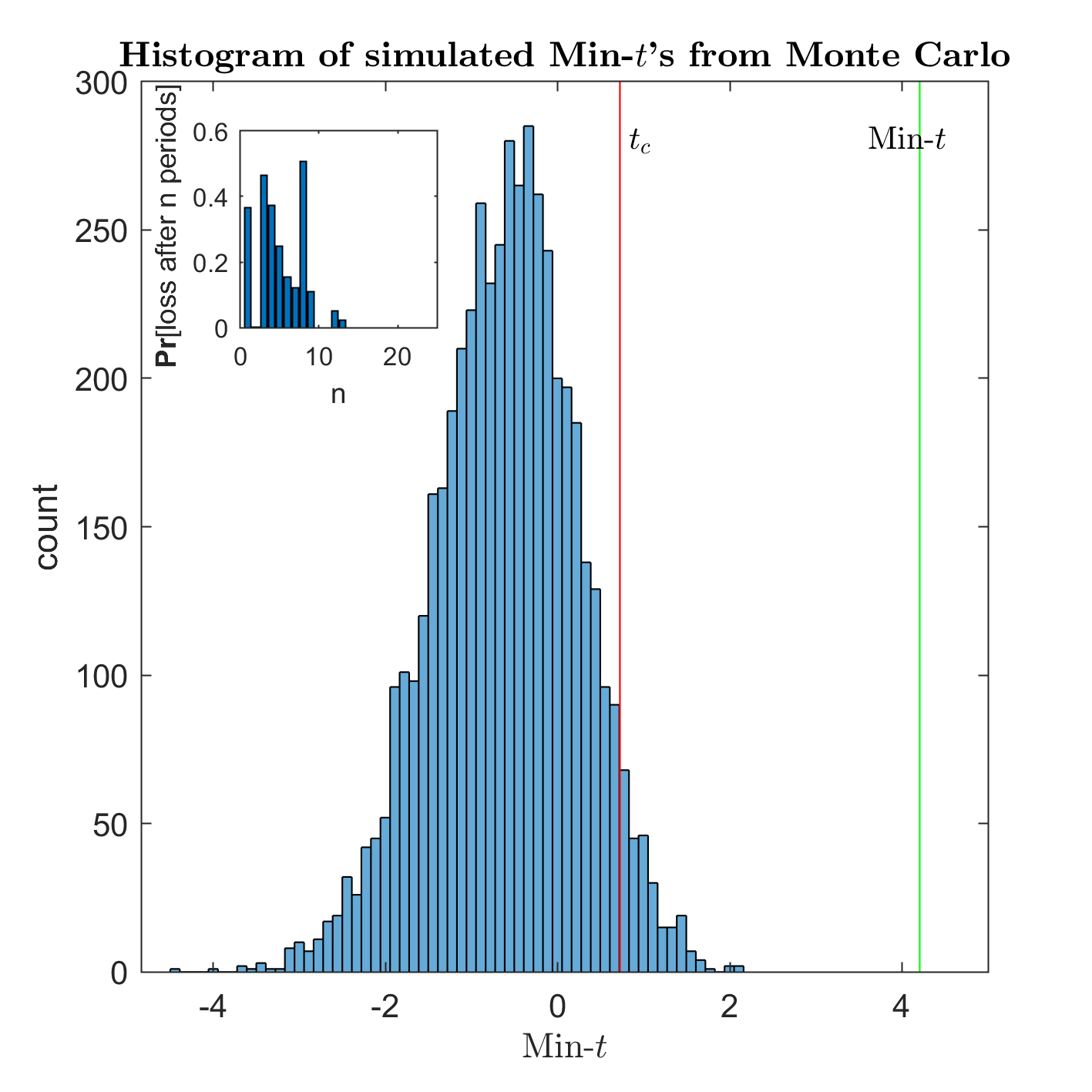}
	\caption{Histogram of the 5000 simulated Min-$t$ statistics resulting from the CM test implemented on the simulated incremental process given in \cref{eq:CMmodel} along with the Min-$t$ statistic (green) for the overall strategy's profit and loss sequence over the 400-day period stretching from the $30^{th}$ trading day until the $430^{th}$ trading day without any account for transactions costs. The figure inset displays the probability of loss for each of the first 30 trading days where we compute the probability of loss of the profit and loss process from the first trading period up to the $n^{th}$ period for each $n= 1, \dots, 25$.}\label{fig:mintMC}
\end{figure}

Next, we implement the CM test for statistical arbitrage on the daily cumulative profits and losses (\textbf{PL}) for the strategy without transaction costs. In order to have a result that is synonymous with \cite{jarrow}, we choose a period of 400 days to test our strategy. We test the realised profits and losses for the 400-day period stretching from the $30^{th}$ trading day until the $430^{th}$ trading day. This is to allow for the algorithm to initiate and leave enough time for majority of the experts to have sufficient data to begin making trading decisions. Having simulated the 5000 different Min-$t$ statistics as in \cref{ssec:statarboutline} step \ref{CMtc} using simulations of the profit process in \cref{eq:CMmodel}, \cref{fig:mintMC} illustrates the histogram of Min-$t$ values. The critical value $t_{c}$ is then computed as the 0.95-quantile of the simulated distribution which refers to a significance level of $\alpha = 5\%$ and is illustrated by the red vertical line. The resulting critical value is $t_c = 0.7263$. The Min-$t$ resulting from the realised incremental profits and losses of the overall strategy is 3.0183 (vertical green line). By \cref{eq:probreject}, we recover a p-value of zero. Thus, we can conclude that there is significant evidence to reject the null of no statistical arbitrage at the 5\% significance level.

In addition to testing for statistical arbitrage, we also report the number of days it takes for the probability of loss of the strategy to decline below 5\% using \cref{eq:probreject} adjusted for the case of the CM model. As discussed in \cref{ssec:statarboutline} step \ref{ploss}, for each $n = 1, \dots, T$, we perform MLE for $\Delta \nu_{1:n}$ to get the parameter estimates. We then substitute these estimates into \cref{eq:probreject} to get an estimate of the probability of loss for the $n^{th}$ period. This is all done in terms of the CM model. The figure inset of \cref{fig:mintMC} illustrates the probability of loss for each of the first 25 trading days, where we compute the probability of loss of the profit and loss process from the first trading period up until the $n^{th}$ period for each $n= 1, \dots, 25$. As is evident from the figure inset, it takes roughly 10 periods for the probability of loss to converge below 5\%. 

\subsubsection{Transaction Costs}

In this section we reproduce the results from above but this time including transaction costs for daily trading as discussed in \cref{ssec:transcosts}.  Once direct and indirect (\cref{eq:cost}) costs have been computed, the idea is to subtract off the transaction cost from the profit and losses of each day and compound the resulting value onto $S_{t-1}$ to get the wealth for period $t$. These daily profit and losses are added to get the cumulative profit and loss $\textbf{PL}$.

\begin{figure*}
	\centering
	\subfigure[Overall cumulative portfolio wealth ($S$) for daily data with transaction costs. The figure inset illustrates the profits and losses (\textbf{PL}) for overall strategy for daily data with transaction costs.]{	\includegraphics[width=7.5cm]{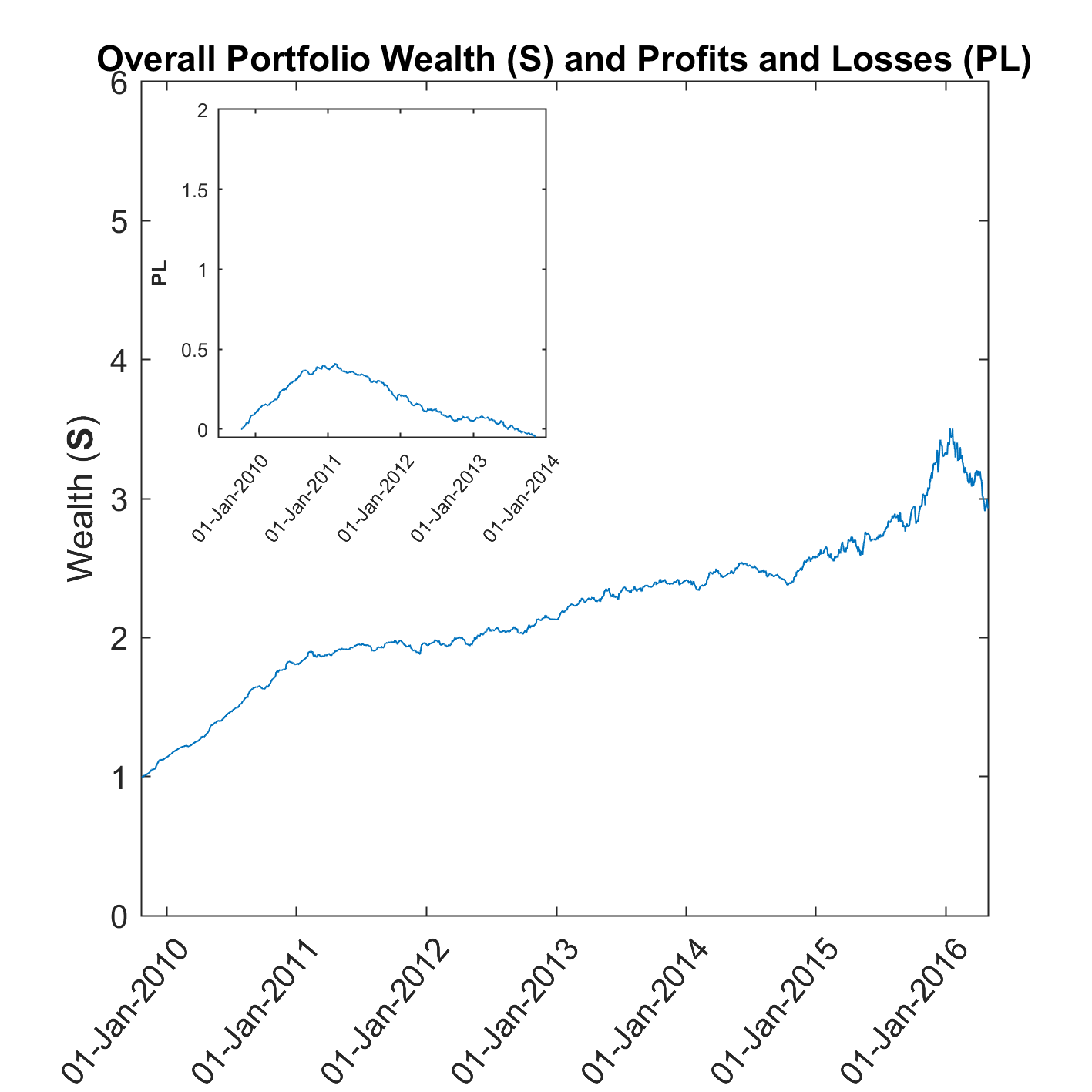}\label{fig:15_strats17_22-Oct-2009-29-Apr-2016_3001_S_TC}}
		\quad
		\subfigure[Histogram of the 5000 simulated Min-$t$ statistics resulting from the CM model and the incremental process given in \cref{eq:CMmodel} along with the Min-$t$ statistic (green) for the overall strategy's profit and loss sequence over the 400 day period stretching from the $30^{th}$ trading day until the $430^{th}$ trading day with transactions costs incorporated. Also illustrated is the critical value at the 5\% significance level (red). The figure inset shows the probability of the overall trading strategy generating a loss for each of the first 400 trading days.]{
		\includegraphics[width=7.5cm]{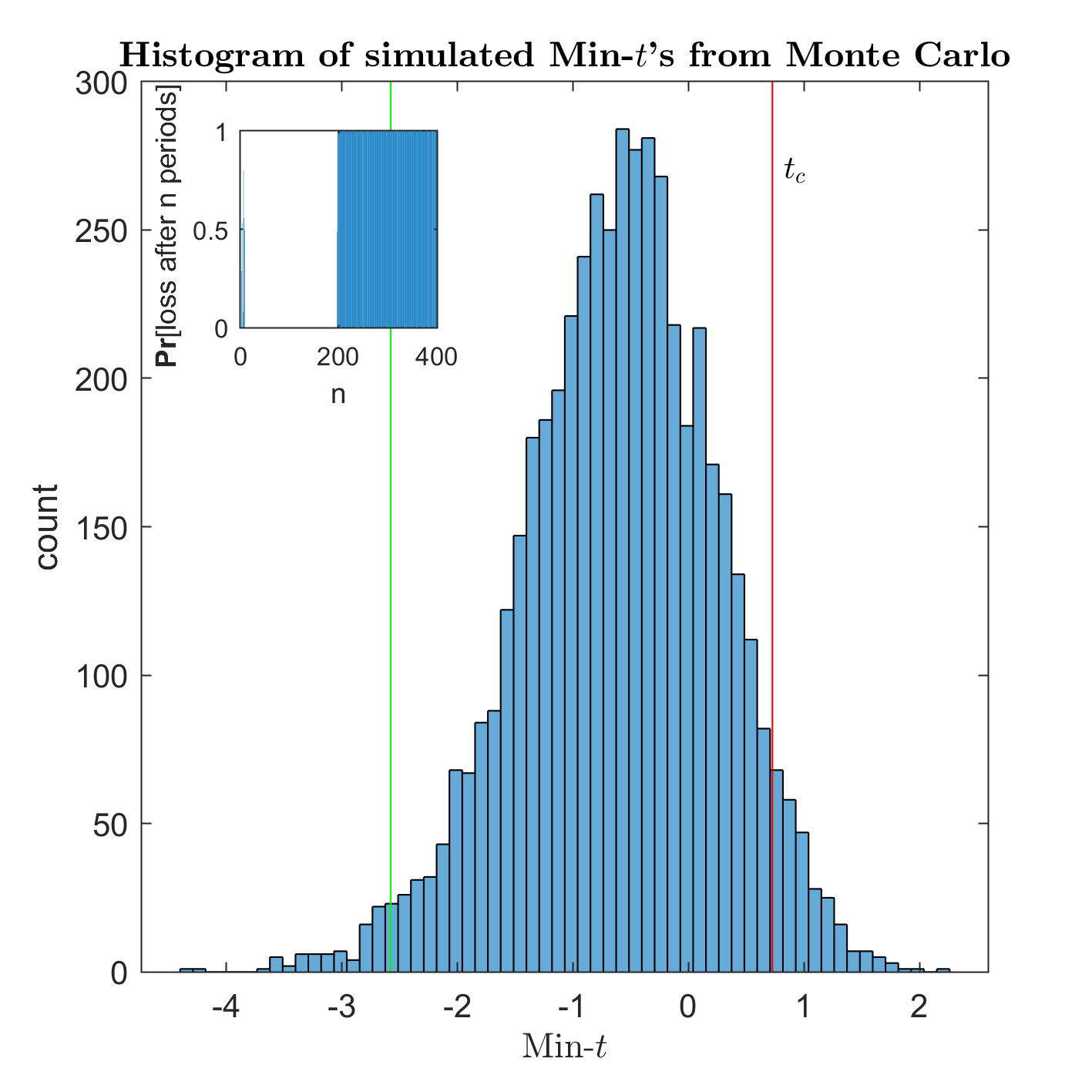}\label{fig:15_strats17_22-Oct-2009-29-Apr-2016_3001_Hist_TC}}
	\caption{The performance of the algorithm (\cref{fig:15_strats17_22-Oct-2009-29-Apr-2016_3001_S_TC}) and the results of the statistical arbitrage test (\cref{fig:15_strats17_22-Oct-2009-29-Apr-2016_3001_Hist_TC}) on daily data witth transaction costs incorporated.}
\end{figure*}

It is clear from the inset of \cref{fig:15_strats17_22-Oct-2009-29-Apr-2016_3001_S_TC}, which illustrates the profits and losses (\textbf{PL}) of the overall strategy less the transaction costs for each period, that consistent losses are incurred when transaction costs are incorporated. Furthermore, there is no evidence to reject the no statistical arbitrage null hypothesis as the Min-$t$ statistic resulting from the overall strategy is well below the critical value at the $95^{th}$ percentile of the histogram as illustrated in \cref{fig:15_strats17_22-Oct-2009-29-Apr-2016_3001_Hist_TC}. In addition to this, although the probability of loss of the strategy with transaction costs included initially converges to zero, it eventually settles on a value of one. This is illustrated in the inset of \cref{fig:15_strats17_22-Oct-2009-29-Apr-2016_3001_Hist_TC}.  

Considering the above evidence contained in \cref{fig:15_strats17_22-Oct-2009-29-Apr-2016_3001_S_TC}, \cref{fig:15_strats17_22-Oct-2009-29-Apr-2016_3001_Hist_TC} and its associated figure inset, the overall strategy does not survive historical back tests in terms of profitability when transaction costs are considered and may not be well suited for an investor utilising daily data whom has a limited time to make adequate profits. This is in agreement with \cite{Schulmeister} in that there is a strong possibility that stock price and volume trends have shifted to higher frequencies than the daily time scale, and resultantly, trading strategies' profits have, over time, diminished on such time scales.

\subsection{Intraday-Daily Data}
%\footnote{See \cref{ssec:sectors_intra} for the list of the 30 stocks along with their Bloomberg ticker symbols}
Below we report the results of the algorithm implementation for a combination of intraday and daily JSE data as discussed in \cref{ssec:int_daily}. We run the algorithm on the OHLCV data of 15 most liquid stocks from a set of 30 of the JSE Top 40. Liquidity is calculated in terms of average daily trade volume for the first 4 days of the period 02-01-2018 to 09-03-2018. The set of 15 stocks is as follows: FSR:SJ, GRT:SJ, SLM:SJ, BGA:SJ, SBK:SJ, WHL:SJ, CFR:SJ, MTN:SJ, DSY:SJ, IMP:SJ, APN:SJ, RMH:SJ, AGL:SJ, VOD:SJ and BIL:SJ. The remaining 40 days' data for the aforementioned period is utilised to run the learning algorithm on. As in the daily data implementation, we again analyse the two cases of trading, with and without transaction costs, which we report in the following two subsections below.  

\subsubsection{No Transaction Costs}

\begin{figure*}
	\centering
	\subfigure[The overall cumulative portfolio wealth (\textbf{S}) for intraday-daily data with no transaction costs. The figure inset illustrates the associated profits and losses.]{
	\includegraphics[width=7cm]{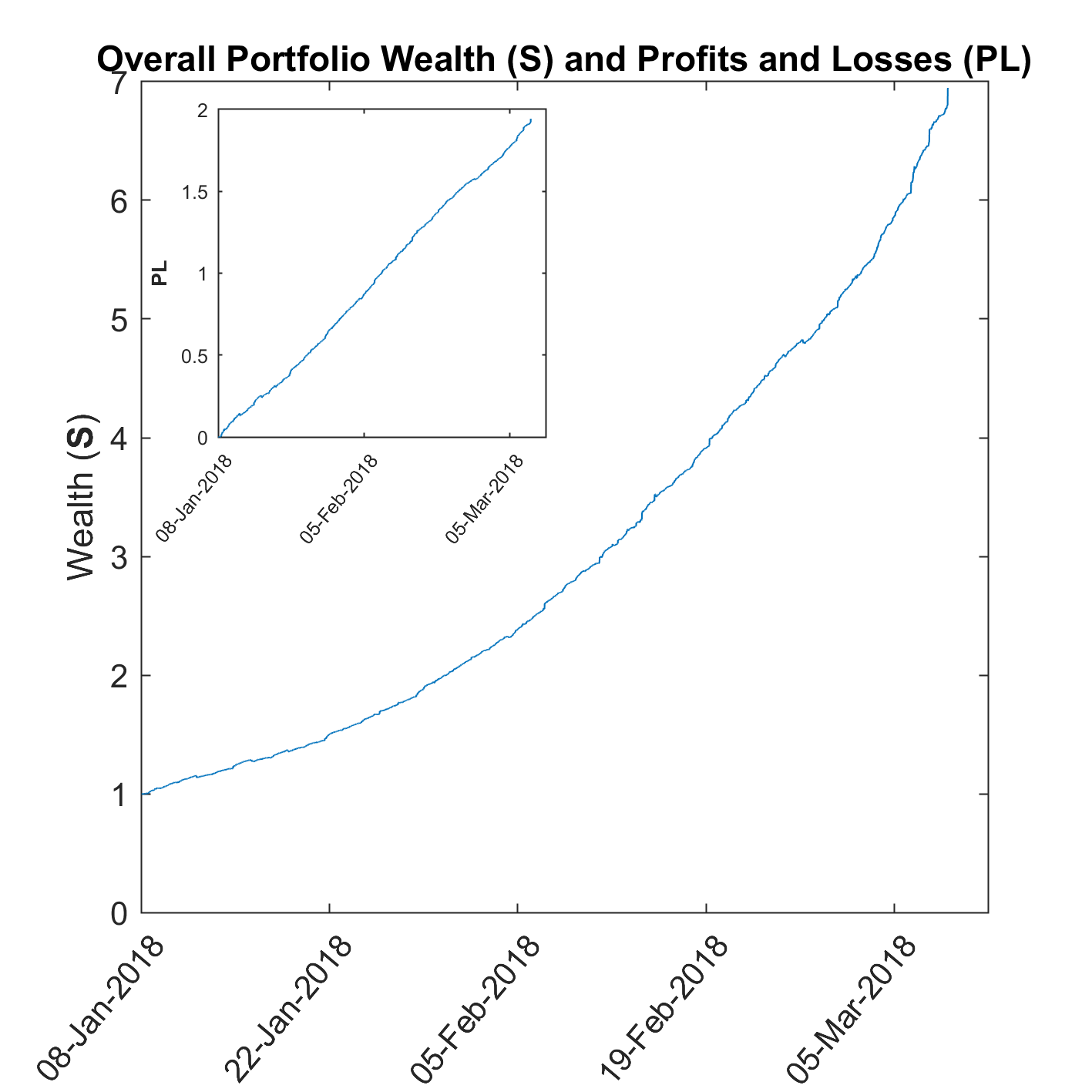}\label{fig:stocks15_strats17_08-Jan-2018-09-Mar-2018_45_S}}
	\quad
	\subfigure[Histogram of the 5000 simulated Min-$t$ statistics resulting from the CM model and the incremental process given in \cref{eq:CMmodel} for the first 400 trading periods for intraday-daily profits and losses without taking into account transaction costs along with the Min-$t$ statistic for the overall strategy (green) and the critical value at the 5\% significance level (red). The figure inset shows the probability of the overall trading strategy generating a loss after $n$ periods for each $n = 5, \dots, 25$ of the intraday-daily profit and loss process (\textbf{PL}) taken from the $5^{th}$ time bar of the second day when active trading commences.]{
	\includegraphics[width=7cm]{15_strats17_08-Jan-2018-09-Mar-2018_45_Hist_pre_print}\label{fig:mintMC_intra}}

	\caption{The performance of the algorithm (\cref{fig:stocks15_strats17_08-Jan-2018-09-Mar-2018_45_S}) and the results of the statistical arbitrage test (\cref{fig:mintMC_intra}) on intraday-daily data without any account for transaction costs.}
\end{figure*}

Without transaction costs, the cumulative wealth achieved by the overall strategy, illustrated in \cref{fig:stocks15_strats17_08-Jan-2018-09-Mar-2018_45_S} evolves similarly to an exponential function over time. The associated profits and losses are displayed in the figure inset of \cref{fig:stocks15_strats17_08-Jan-2018-09-Mar-2018_45_S}. Incremental profits and losses are obviously a lot smaller compared to the daily data cases resulting in a much smoother function in comparison to the daily data case (\cref{fig:20_strats16_22-Oct-2009-29-Apr-2016_3001_S}).

\cref{tab:stats_wealth_intra} is the intraday-daily analogue of \cref{tab:stats_wealth}. In this case, the exponential moving crossover strategy (EMA X-over) produces the expert with the greatest wealth and acceleration (ACC) the expert with the least terminal wealth. Exponential moving crossover also produces experts with the highest variation in terminal wealth's. Price rate of change (PROC) comfortably provides the best mean ranking experts among all experts among all other strategies, however, Z-BCRP produces experts with highest mean terminal wealth. 

Again, as for the daily data case, we implement a test for statistical arbitrage for intraday-daily trading without transaction costs for 400 trading periods starting from the $6^{th}$ time bar of the 2nd trading day\footnote{This corresponds to the trading period within which the very first trading decisions are made.} using the intraday-daily profit and loss sequence (\textbf{PL}). \cref{fig:mintMC_intra} illustrates the histogram of simulated Min-$t$ values with the 0.95-percentile of the simulated distribution representing the critical value $t_{c}$ (red) and the Min-$t$ (green) resulting from the incremental profits and losses of the overall strategy resulting from the learning algorithm. The resulting critical value is 0.7234 and the Min-$t$ value is 4.2052. Thus, there is strong evidence to reject the null hypothesis of no statistical arbitrage as the resulting p-value is identical to zero. 
\LTcapwidth=14cm
\begin{longtable}[t]{p{4cm}p{3cm}p{2cm}p{1.5cm}p{1cm}}
	\caption{Group summary statistics of the overall rankings of experts grouped by their underlying strategy ($\boldsymbol{\omega}(i)$ where $i=1, \dots, 17$) for intraday-daily trading. In brackets are the mean overall ranking of experts utilising each strategy.}\label{tab:stats_wealth_intra}\\
	\hline
	\rule{0pt}{3ex}
	Strategy & Mean (mean rank) & St. Dev. & Min  & Max  \\
	\hline
	\rule{0pt}{3ex}	
	\noindent EMA X-over&	1.0024 (662.7639)	&0.0094	&0.9801&	1.0375\\
	Ichimoku Kijun Sen &	0.9989 (710.3750)&	0.0085&	0.9663&	1.0303\\
	MACD & 	0.9995 (684.8704) &	0.0067&	0.9720	&1.0202\\
	Moving Ave X-over  &	1.0012 (708.7824)&	0.0058&	0.9766&	1.0204\\
	ACC & 	0.9953 (831.3333) &	0.0079&	0.9646&	1.0048\\
	BOLL & 	0.9974 (712.9722)&	0.0069&	0.9787&	1.0089\\
	Fast Stochastic& 	0.9991 (711.4167)&	0.0040&	0.9871&	1.0085\\
	MARSI & 	0.9973 (736.2500)&	0.0062&	0.9824&	1.0094\\
	MOM & 	0.9982 (723.1389)&	0.0087&	0.9700&	1.0082\\
	Online Anti-Z-BCRP& 0.9980 (597.3056)&	0.0062&	0.9828&	1.0103\\
	Online Z-Anticor & 	1.0015 (655.7778)&	0.0058&	0.9896&	1.0180\\
	Online Z-BCRP & 	1.0031 (566.5833)&	0.0069&	0.9898&	1.0149\\
	PROC & 	0.9980 (445.1389)&	0.0064&	0.9814&	1.0140\\
	RSI &	0.9997 (535.5833)&	0.0065&	0.9861&	1.0171\\
	SAR &	0.9945 (499.7222)&	0.0053&	0.9790&	1.0005\\
	Slow Stochastic &	1.0007 (508.5278)&	0.0048&	0.9927&	1.0173\\
	Williams \%R &	1.0020 (536)&	0.0034&	0.9957&	1.0133\\
	\hline
\end{longtable}

The figure inset of \cref{fig:mintMC_intra} illustrates the probability of loss for each of the first 25 periods of the 400 periods as discussed in the above paragraph. It takes roughly an hour (13 periods) for the probability of loss to converge to zero.

\subsubsection{Transaction Costs}

\begin{figure}%[t]
	\centering
	\subfigure[The overall cumulative portfolio wealth ($\textbf{S}$) for intraday-daily data with transaction costs. The figure inset illustrates the associated profits and losses.]{	\includegraphics[width=7.5cm]{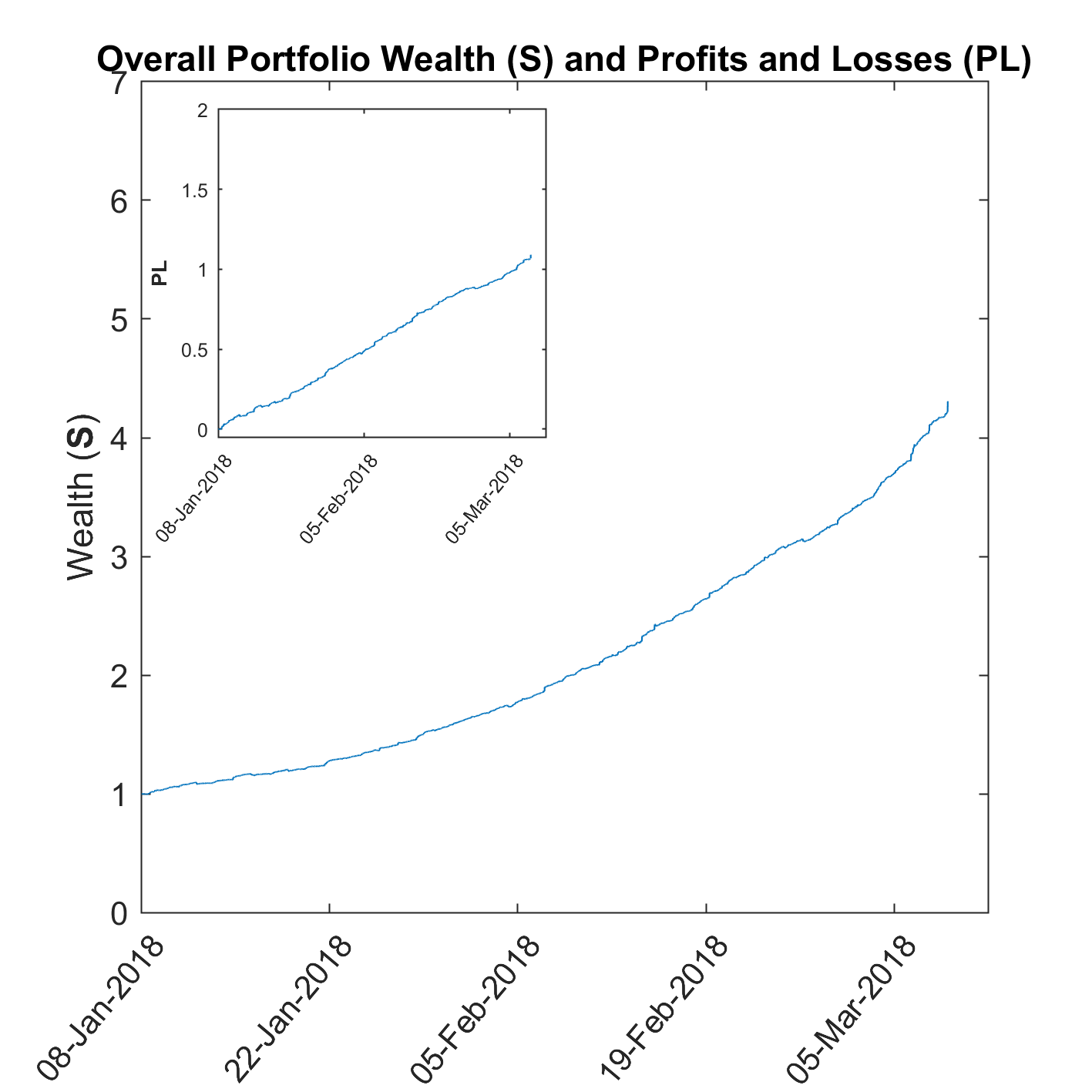}\label{fig:stocks15_strats17_08-Jan-2018-09-Mar-2018_45_S_TC}}
	\quad
	\subfigure[Histogram of the 5000 simulated Min-$t$ statistics resulting from the CM model and the incremental process given in \cref{eq:CMmodel} for the first 400 trading periods for intraday-daily profit and losses less transaction costs along with the Min-$t$ statistic for the overall strategy (green) and the critical value at the 5\% significance level (red).]{
		\includegraphics[width=7.5cm]{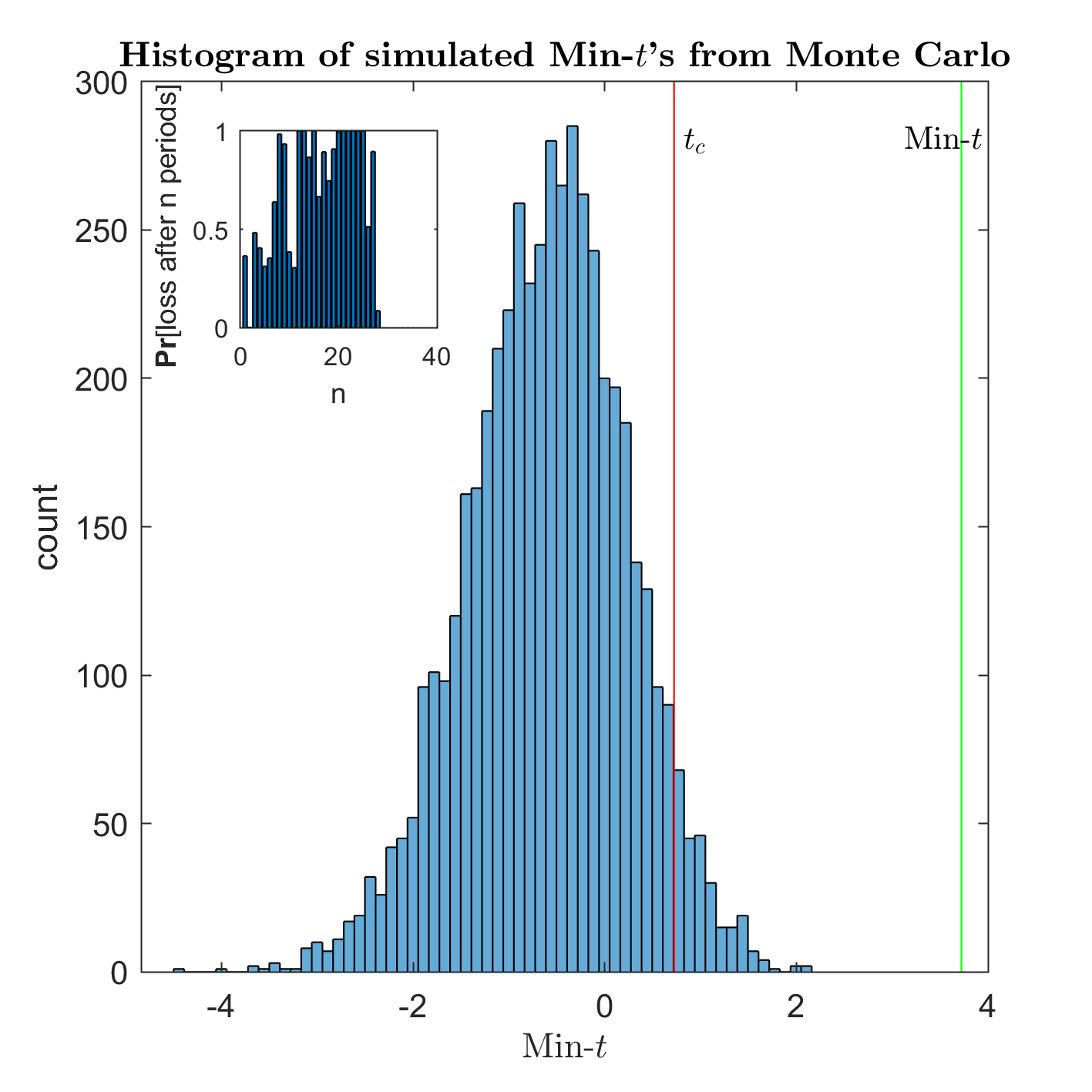}\label{fig:15_strats17_08-Jan-2018-09-Mar-2018_45_Hist_TC} }
	\caption{The performance of the algorithm (\cref{fig:stocks15_strats17_08-Jan-2018-09-Mar-2018_45_S_TC}) and the results of the statistical arbitrage test (\cref{fig:15_strats17_08-Jan-2018-09-Mar-2018_45_Hist_TC}) on intraday-daily data with transaction costs incorporated.}
\end{figure}

We now report the results of the algorithm run on the same intraday-daily data as in the subsection above but this time with transaction costs incorporated (see \cref{ssec:transcosts}). \cref{fig:stocks15_strats17_08-Jan-2018-09-Mar-2018_45_S_TC} and the figure inset illustrate the overall cumulative portfolio wealth ($\textbf{S}$) and profits and losses (\textbf{PL}) respectively for intraday-daily trading with transaction costs. For comparative reasons, the axes are set to be equivalent to those illustrated in the case of no transaction costs (\cref{fig:stocks15_strats17_08-Jan-2018-09-Mar-2018_45_S_TC} and the figure inset). Surprisingly, even with a total daily trading cost (direct and indirect) of roughly 130bps, which is a fairly conservative approach, the algorithm is able to make satisfactory returns, which is in contrast to the daily trading case (\cref{fig:15_strats17_22-Oct-2009-29-Apr-2016_3001_S_TC}). Furthermore, \cref{fig:15_strats17_08-Jan-2018-09-Mar-2018_45_Hist_TC} provides significant evidence to reject the no statistical arbitrage null hypothesis and returns a Min-$t$ statistic almost identical (4.32 in the transaction costs case compared to 3.87) to that of the case of no transaction costs (\cref{fig:mintMC_intra}). Even more comforting, is the fact that even when transaction costs are considered, the probability of loss per trading period converges to zero, albeit slightly slower (roughly 2 hours or 31 trading periods) than the case of no transaction costs (roughly 1 hour or 13 trading periods, as illustrated in the inset of \cref{fig:mintMC_intra}). 

The above results for intraday-daily trading are in complete contrast to the case of daily trading with transaction costs, whereby the no statistical arbitrage null could not be rejected, the probability of loss did not converge to zero and remain there, and trading profits steadily declined over the trading horizon. This suggests that the proposed algorithm may be much better suited to trading at higher frequencies. This is not surprising and is in complete agreement with \cite{Schulmeister} who argues that the profitability of technical trading strategies had declined over from 1960, before becoming unprofitable from the 1990's. A substantial set of technical trading strategies are then implemented on 30-minute data and the evidence suggests that such strategies returned adequate profits between 1983 and 2007 however the profits declined slightly between 2000 and 2007 compared to the 1980's and 1990's. This suggests that markets may have become more efficient and even the possibility that stock price and volume trends have shifted to even higher frequencies than 30 minutes (\cite{Schulmeister}). This supports the choice to trade the algorithm proposed in this paper on at least 5-minute OHLCV data and reinforces our conclusion that ultimately, the most desirable implementation of the algorithm would be in volume-time, which is best suited for high frequency trading.

\section{Conclusion} \label{sec:conc}

We have developed a learning algorithm built from a base of technical trading strategies for the purpose of trading equities on the JSE that is able to provide favourable returns when ignoring transaction costs, under both daily and intraday trading conditions. The returns are reduced when transaction costs are considered in the daily setting, however there is sufficient evidence to suggest that the proposed algorithm is really well suited to intraday trading. 

This is reinforced by the fact that there exists meaningful evidence to reject a carefully defined null hypothesis of no statistical arbitrage in the overall trading strategy even when a reasonably aggressive view is taken on intraday trading costs. We are also able to show that it in both the daily and intraday-daily data implementations that the probability of loss declines below 5\% relatively quickly which strongly suggests that the algorithm is well suited for a trader whose preference or requirement is to make adequate returns in the short-run. It may well be that the statistical arbitrages we have identified intraday are artefacts’ from ``price distorters'' (\cite{MS2017}) rather than legitimate mispricing in the sense of majority views relative to a trading minority and hence cannot be easily traded out of profit. This suggests that it can be important to try unpack the difference between the structural mispricing’s relative to statistical arbitrages---that is outside of the scope of the current work and cannot be determined using the tests implemented in this work.

The superior performance of the algorithm for intraday trading is in agreement with \cite{Schulmeister}, who concluded that while the daily profitability of a large set of technical trading strategies has steadily declined since 1960 and has been unprofitable since the onset of the 1990's, trading the same strategies on 30-minute (intraday) data between 1983 and 2007 has produced decent average gross returns. However, such returns have slowly declined since the early 2000's. In conclusion, the proposed algorithm is much better suited to trading at higher frequencies; but we are also aware that over time tradings strategies that are not structural in nature are slowly arbitraged away through over-crowding.

We are also cognisant of the fact that intraday trading will require a large component of accumulated trading profits to finance frictions, concretely to fund direct, indirect and business model costs (\cite{loonat}). For this reason, we are careful to remain sceptical with this class of algorithms long-run performance when trading with real money in a live trading environment for profit. The current design of the algorithm is not yet ready to be traded on live market data, however with some effort it is easily transferable to such use cases given the sequential nature of the algorithm and its inherent ability to receive and adapt to new incoming data while making appropriate trading decisions based on the new data. Concretely, the algorithm should be deployed in the context of volume-time trading rather than the calendar time context considered in this work. 

Possible future work includes implementing the algorithm in volume-time, which will be best suited for dealing with a high frequency implementation of the proposed algorithm, given the intermittent nature of order-flow. We also propose replacing the learning algorithm with an online (adaptive) neural network that has the ability to predict optimal holding times of stocks. Another interesting line of work that has been considered is to model the population of trading experts as competing in a predator-prey environment (\cite{Farmerpredprey,johnson}). This was an initial key motivation for the research project, to find which collections of technical trading strategies can be grouped collectively and how these would interact with each other. This can include using cluster analysis to group, or separate trading experts, based on their similarities and dissimilarities, and hence make appropriate inferences regarding their interactions and behaviours at the level of collective and emergent dynamics. This can in turn be used for cluster based approaches for portfolio control. 
	
\subsection*{Acknowledgements}
NM and TG would like to thank the Statistical Finance Research Group in the Department of Statistical Sciences for various useful discussions relating to the work. In particular, we would like to thank Etienne Pienaar, Lionel Yelibi and Duncan Saffy. We thank Michael Gant for his help with developing some of the strategies and for numerous valuable discussions with regards to the statistical arbitrage test. 

\subsection*{Funding}
TG would like to thank UCT FRC for funding (UCT fund 459282). 

\subsection*{Supplemental material}\label{ssec:supplemental}
Please access the supplemental material at \cite{git}.

\appendices

\newcounter{tblEqCounter} %create a counter
\setcounter{tblEqCounter}{\theequation} %at the start of the table, set the counter to equation numbering
\section{Technical Indicators and Trading Rules}\label{ssec:inds}
We follow \cite{creamer,QuantStrat} in introducing and describing some of the more popular technical analysis indicators as well as a few others that are widely available. We also provide some trading rules which use technical indicators to generate buy, sell and hold signals.
\LTcapwidth=13cm
\renewcommand*{\arraystretch}{1}

{\setlength{\tabcolsep}{10pt}
\begin{small}
\begin{longtable}{p{0.13\textwidth}p{0.25\textwidth}p{0.6\textwidth}}
	\caption{The set of trading indicators utilised by the trading rules described in Table \ref{tab:rules} along with their descriptions and calculation details.}\label{tab:inds}\\
			\toprule
			Indicator & Description & Calculation\\
			\toprule

$\mbox{SMA}^{c}_{t}(n)$ & The \textit{Simple Moving Average} (SMA) is the mean of the closing prices over the last $n$ trading days. The smaller the value of $n$, the closer the moving average will fit to the price data. & 
\begin{equation}
			\mbox{SMA}_{t}(\textbf{P}^{c}, n) = \frac{1}{n}\sum_{i=0}^{n-1}P_{t-i}^{c}\end{equation}\\

$\mbox{EMA}^{c}_{t}(n)$ &  The {\it Exponential Moving Average} (EMA) uses today's close price, yesterday's moving average value and a smoothing factor ($\alpha$). The smoothing factor determines how quickly the exponential moving average responds to current market prices (\cite{QuantStrat}). & 
\begin{eqnarray}
&\mbox{EMA}_{t}(\textbf{P}^{c}, n) = \alpha P_{t}^{c} + (1-\alpha)\mbox{EMA}_{t-1}(P_{t-1}^{c}, n) \\ 
&\mbox{where} ~\alpha = \frac{2}{n+1}, ~\mbox{and}~
\mbox{EMA}_{0}(P_{0}^{c}, n) = \mbox{SMA}(\textbf{P}^{c}, n)
\end{eqnarray} \\

\mbox{HH}(n) & The {\it Highest High} (HH) is the greatest high price in the last $n$ periods and is determined from the vector $\boldsymbol{P^{h}_{n}}$  of the high-prices of the last $n$ periods. & 
Given the high prices of last $n$ periods:
\begin{equation} 
\boldsymbol{P^{h}_{n}} = (P^{h}_{t-n}, P^{h}_{t-n+1}, P^{h}_{t-n+2},\dots  , P^{h}_{t})
\end{equation}
to find:
\begin{equation}
\mbox{HH}(n) = \mbox{max}(\boldsymbol{P^{h}_{n}}) .
\end{equation}
\\
			
\mbox{LL}(n) & {\it Lowest Low} (LL) is the smallest low price in the last $n$ periods and is found from the vector $\boldsymbol{P^{l}_{n}}$ of low prices in the last $n$ periods.  & Givne the low prices of the last $n$ periods:
\begin{equation}
\boldsymbol{P^{l}_{n}} = (P^{l}_{t-n}, P^{l}_{t-n+1}, P^{l}_{t-n+2}, \dots, P^{l}_{t})
\end{equation}
to find:
\begin{equation}
\mbox{LL}(n) = \mbox{min}(\boldsymbol{P^{l}_{n}}).
\end{equation}\\
			
			$\mbox{IKH}(n_1,n_2,n_3)$ & The {\it Ichimoku Kinko Hyo} (IKH) (at a glance equilibrium chart) system consists of five lines and the Kumo (cloud).\footnote{See \cite{linton2010cloud,kumotrader,ichimoku101}} The five lines all work in concert to produce the end result. The size of the Kumo is an indication of the current market volatility, where a wider Kumo is a more volatile market. Typical input parameters: $n_{1}=7$, $n_{2}=22$, and $n_{3}=44$. Here we keep $n_{1}$ fixed at 7 but vary the other two parameters. &  
				Tenkan-sen (Conversion Line): 
\begin{equation}
\mbox{HH}(n_{1}) + \mbox{LL}(n_{1}))/2 
\end{equation}
				Kijun-sen (Base Line): 
\begin{equation}
\mbox{HH}(n_{2}) + \mbox{LL}(n_{2}))/2 
\end{equation}
				Chikou Span (Lagging Span): 
\begin{equation}
\mbox{Close plotted}~n_{2}~ \mbox{days in the past}
\end{equation}
				Senkou Span A (Leading Span A):
\begin{equation}
(\mbox{Conversion Line} + \mbox{Base Line})/2
\end{equation}
				Senkou Span B (Leading Span B):
\begin{equation}
\mbox{HH}(n_{3}) + \mbox{LL}(n_{3}))/2 
\end{equation}
				\makecell[lt]{Kumo (Cloud): Area between the Leading Span A\\ and the Leading Span B from the Cloud} \\

$\mbox{MOM}_{t}(n)$  &  {\it Momentum} (MOM) gives the change in the closing price over the past $n$ periods.  & 
\begin{equation}
\mbox{MOM}_{t}(n) = P^{c}_{t} - P^{c}_{t-n}
\end{equation} \\

$\mbox{ACC}_{t}(n)$ & {\it Acceleration}(ACC) measures the change in momentum between two consecutive periods $t$ and $t-1$  & 
\begin{equation}
\mbox{ACC}_{t}(n) = \mbox{MOM}_{t}(n) - \mbox{MOM}_{t-1}(n)
\end{equation}\\
		    	
		    	$\mbox{MACD}_{t}(n_{1},n_{2})$ & The {\it Moving Average Convergence/Divergence} (MACD) oscillator attempts to determine whether traders are accumulating stocks or distributing stocks. It is calculated by computing the difference between a short-term and a long-term moving average. A signal line is computed by taking an EMA of the MACD and determines the instances to buy (over-sold) and sell (over-bought) when used in conjunction with the MACD\footnote{See \cite{PHDRechenthin}}. & 
Long-term EWM:
\begin{equation}
\mbox{EMA}^L_{t}=\mbox{EMA}_{t}(P^{c},n_{2}) 
\end{equation}
Short-term EWM:
\begin{equation}
\mbox{EMA}^S_{t}= \mbox{EMA}_{t}(P^{c},n_{1})
\end{equation}
Moving Average Convergence/Divergence:
\begin{equation}
\mbox{MACD}_{t}(n_{1}, n_{2}) = \mbox{EMA}^S_{t} - \mbox{EMA}^L_{t}
\end{equation}
The ``Signal Line" (SL):
\begin{equation}
\mbox{SL}_{t}(n_{1}, n_{2}, n_{3}) = \mbox{EMA}_{t}(\mbox{MACD}_{t}(n_{2},n_{1}),n_{3}) 
\end{equation}
The ``MACD Signal" (MACDS):
\begin{equation}
\mbox{MACDS}_{t}(n_{1}, n_{2}, n_{3}) = \mbox{MACD}_{t}(n_{1}, n_{2}) -
\mbox{SL}_{t}(n_{1}, n_{2}, n_{3})
\end{equation}
\\	
\makecell[lt]{$\mbox{Fast\%K}_{t}(\textbf{n})$\\ ~~~~and\\ $\mbox{Fast\%D}_{t}(n)$} & {\it Fast Stochastic Oscillator} shows the location of the closing price relative to the high-low range, expressed as a percentage, over a given number of periods as specified by a look-back parameter. & 
\begin{equation}
\mbox{Fast\%K}_{t}(n) = \frac{P_{t}^{c} - \mbox{LL}(n)}{\mbox{HH}(n)-\mbox{LL}(n)}
\end{equation}
\begin{equation}
\mbox{Fast\%D}_{t}(n) = \mbox{SMA}_{t}(\mbox{Fast\%K}_{t}(n),3)
\end{equation}\\
\\
\makecell[lt]{$\mbox{Slow\%K}_{t}(n)$ \\ ~~~~and\\ $\mbox{Slow\%D}_{t}(n)$} & The {\it Slow Stochastic Oscillator} is very similar to the fast stochastic indicator and is in fact just a moving average of the fast stochastic indicator. & 
\begin{equation}
\mbox{Slow\%K}_{t}(n) = \mbox{SMA}_{t}(\mbox{Fast\%K}_{t}(n),3)
\end{equation}
\begin{equation}
\mbox{Slow\%D}_{t}(n) = \mbox{SMA}_{t}(\mbox{Slow\%K}_{t}(n),3)
\end{equation}\\
\\
\makecell[lt]{$\mbox{RSI}_{t}(n)$} & {\it Relative Strength Index} (RSI) compares the periods that stock prices finish up (closing price higher than the previous period) against those periods that stock prices finish down (closing price lower than the previous period).\footnote{See \cite{creamer}} & 
\begin{equation}
\mbox{RSI}_{t}(n) = 100-\frac{100}{1+\frac{\mbox{SMA}_{t}(\boldsymbol{P}_{n}^{\text{up}},n_{1})}{\mbox{SMA}_{t}(\boldsymbol{P}_{n}^{\text{dwn}},n_{1})}}
\end{equation}
where finishing up (down) are:			
\begin{equation}
P_{t}^{\text{up}} = \begin{cases} P_{t}^{c} & \text{if}\ P_{t-1}^{c} < P_{t}^{c} \\
\text{NaN} & \text{otherwise}\end{cases}
\end{equation}
\begin{equation}
P_{t}^{\text{down}} = \begin{cases} P_{t}^{c} & \text{if}\ P_{t-1}^{c} > P_{t}^{c} \\ \text{NaN} & \text{otherwise}\end{cases}
\end{equation}
to find the vector of up and down finishing cases:
\begin{equation}
\boldsymbol{P}_{n}^{\text{up}} = (P_{t-n}^{\text{up}}, P_{t-n+1}^{\text{up}}, \dots, P_{t}^{\text{up}})
\end{equation}
\begin{equation}
\boldsymbol{P}_{n}^{\text{down}} = (P_{t-n}^{\text{down}}, P_{t-n+1}^{\text{down}}, \dots, P_{t}^{\text{down}})
\end{equation}\\
$\mbox{MARSI}_{t}(n_{1},n_{2})$ & {\it Moving Average Relative Strength Index} (MARSI) is an indicator that smooths out the action of RSI indicator.\footnote{See \cite{Marsi}} MARSI is calculated by simply taking an $n_{2}$-period SMA of the RSI indicator.  & 
\begin{equation}
\mbox{MARSI}_{t}(n_{1},n_{2}) = \mbox{SMA}(\mbox{RSI}_{t}(n_{1}),n_{2})
\end{equation} \\
				 
$\mbox{Boll}^{m}_{t}(n)$ & {\it Bollinger}(Boll) bands uses a SMA ($\mbox{Boll}^{m}_{t}(n)$) as it's reference point (known as the median band) with regards to the upper and lower Bollinger bands denoted by $\mbox{Boll}^{u}_{t}(n)$ and $\mbox{Boll}^{d}_{t}(n)$ respectively and are calculated as functions of standard deviations ($s$). 
& Median band:
\begin{equation}
\mbox{Boll}^{m}_{t}(n) = \mbox{SMA}^{c}_{t}(n)
\end{equation}
Upper band: 
\begin{equation}
\mbox{Boll}^{u}_{t}(n)=\mbox{Boll}^{m}_{t}(n) + s \sigma^{2}_{t}(n)
\end{equation}
Lower band: 
\begin{equation}
\mbox{Boll}^{d}_{t}(n)=\mbox{Boll}^{m}_{t}(n) - s \sigma^{2}_{t}(n)
\end{equation}
Here $s$ is chosen to be 2.\\

$\mbox{PROC}_{t}(n)$ & The rate of change of the time series of closing prices $P_{t}^{c}$ over the last $n$ periods expressed as a percentage. &	
\begin{equation}
\mbox{PROC}_{t}(n) = 100\cdot\frac{P_{t}^{c} - P_{t-n}^{c}}{P_{t-n}^{c}}
\end{equation}\\
				
$\mbox{Will}_{t}(n)$ & {\it Williams Percent Range} (Williams \%R) is calculated similarly to the fast stochastic oscillator and shows the level of the close relative to the highest high in the last $n$ periods. & 
\begin{equation}
\mbox{Will}_{t}(n) = \frac{\mbox{HH}(n) - P_{t}^{c}}{\mbox{HH}(n) - \mbox{LL}(n)} \cdot (-100)
\end{equation}\\
				
				$\mbox{SAR}(n)$ & {\it Parabolic Stop and Reverse} (SAR), developed by J. Wells Wilder, is a trend indicator formed by a parabolic line made up of dots at each time step (\cite{SAR}). The dots are formed using the most recent Extreme Price and an acceleration factor (AF), 0.02, which increases each time a new Extreme Price (EP) is reached. The AF has a maximum value of 0.2 to prevent it from getting too large. Extreme Price represents the highest (lowest) value reached by the price in the current up-trend (down-trend). The acceleration factor determines where in relation to the price the parabolic line will appear by increasing by the value of the AF each time a new EP is observed and thus affects the rate of change of the Parabolic SAR.  & 
Calculating the SAR indicator:
\begin{enumerate}
\item \makecell[lt]{{\bf Initialise}:  Set initial trend to 1 (up-trend),\\ EP to zero, $AF_{0}$ to 0.02, $\mbox{SAR}_{0}$ to the closing\\ price at time zero ($P^{c}_{0}$), Last High (LH) to\\ high price at time zero ($P^{h}_{0}$) and Last Low\\ (LL) to the low price at time zero ($P^{l}_{0}$))}
\item \makecell[lt]{{\bf Level Update}: update EP, LH, LL and AF\\ based on the current high in relation to the\\ LH (up-trend), or where the current low is in\\ relation to the LL (down-trend)}
\item \makecell[lt]{{\bf Time Update}: update time $t+1$ SAR value,\\ $\mbox{SAR}_{t+1}$, using equation \eqref{eq:sar} for the Parabolic\\ SAR for time $t+1$ as calculated using the\\ previous value at time $t$:}
\begin{equation}
 \mbox{SAR}_{t+1}=\mbox{SAR}_{t}+\alpha (\mbox{EP}-\mbox{SAR}_{t})
\end{equation}
\item \makecell[lt]{{\bf Trend Update}: modify the $\mbox{SAR}_{t+1}$ value,\\ AF, EP, LL, LH and the trend based on the\\ trend and it's value in relation to the current\\ low $P^{l}_{t}$ and current high $P^{h}_{0}$} 
\item \makecell[lt]{{\bf Iterate}: go to next time period and return to\\ step 2}
\end{enumerate}\\
\botrule
\end{longtable}
\end{small}
}

{\setlength{\tabcolsep}{10pt}
\begin{small}
\begin{longtable}{p{0.15\textwidth}p{0.08\textwidth}p{0.77\textwidth}}
		\caption{The set of trading rules implemented in the study and the associated conditions under which buy, sell and hold decisions are made under each rule.}\label{tab:rules}
		\toprule
		Trading Rule & Decision & Condition

		\toprule
			Moving Average Crossover & \makecell[lt]{\noindent\textit{Buy} \\[6.5mm] \textit{Sell} \\[7mm] \textit{Hold}}   & \makecell[lt]{$SMA_{t-1}(n_{1}) < SMA_{t-1}(n_{2})$ \& $SMA_{t}(n_{1}) >= SMA_{t}(n_{2})$\\[1mm] where $n_{1} < n_{2}$ \\[1mm] $SMA_{t-1}(n_{1}) > SMA_{t-1}(n_{2})$ \& $SMA_{t}(n_{1}) <= SMA_{t}(n_{2})$\\[1mm] where $n_{1} < n_{2}$ \\[1mm] otherwise}\\[10pt]
		
			Exponential Moving Average Crossover & \makecell[lt]{\noindent\textit{Buy} \\[7mm] \textit{Sell} \\[1mm] \textit{Hold}}  & \makecell[lt]{$EMA_{t-1}(n_{1}) < EMA_{t-1}(n_{2})$ \& $EMA_{t}(n_{1}) >= EMA_{t}(n_{2})$\\[1mm] where $n_{1} < n_{2}$ \\[1mm] $EMA_{t-1}(n_{1}) > EMA_{t-1}(n_{2})$ \& $EMA_{t}(n_{1}) <= EMA_{t}(n_{2})$ \\[1mm] otherwise}\\
		
			Ichimoku Kijun Sen Cross & \makecell[lt]{\noindent\textit{Buy} \\[1mm] \textit{Sell} \\[1mm] \textit{Hold}} & \makecell[lt]{Kijun Sen crosses the closing price curve from the bottom up\\[1mm] Kijun Sen crosses the closing price curve from the top down \\[1mm] otherwise}\\
			
			Momentum & \makecell[lt]{\noindent\textit{Buy} \\[5mm] \textit{Sell} \\[4.5mm] \textit{Hold}} &  \makecell[lt]{$\mbox{MOM}_{t-1}(n) \leq 	\mbox{EMA}_{t}(\mbox{MOM}_{t}(n),\lambda)$
			\& \\$MOM_{t}(n) > EMA_{t}(MOM_{t}(n),\lambda)$\\[1mm]
			$MOM_{t-1}(n) \geq \mbox{EMA}_{t}(\mbox{MOM}_{t}(n),\lambda)$
			\& \\$\mbox{MOM}_{t}(n) < \mbox{EMA}_{t}(\mbox{MOM}_{t}(n),\lambda)$\\[1mm]
		 otherwise}\\
	 
		 	Acceleration & 	 \makecell[lt]{\noindent\textit{Buy} \\[1mm] \textit{Sell} \\[1mm] \textit{Hold}} & \makecell[lt]{$\mbox{ACCEL}_{t-1}(n) +1 \leq 0$
		 	\& $\mbox{ACCEL}_{t}(n) +1 > 0$\\[1mm] 
	 		 $\mbox{ACCEL}_{t-1}(n) + 1 \geq 0$
		 	\& $\mbox{ACCEL}_{t}(n) + 1 < 0$\\[1mm] 
	 		 otherwise} \\
 	 
 	 	MACD &  \makecell[lt]{\noindent\textit{Buy} \\[5mm] \textit{Sell} \\[4.5mm] \textit{Hold}} & \makecell[lt]{$\mbox{MACD}_{t-1}(n_{2},n_{1}) \leq \mbox{MACDS}_{t}(n_{2},n_{1}, n_{3})$
 		 \& \\$\mbox{MACD}_{t}(n_{2},n_{1}) > \mbox{MACDS}_{t}(n_{2},n_{1} n_{3})$ \\[1mm]
		 $\mbox{MACD}_{t-1}(n_{2},n_{1}) \geq \mbox{MACDS}_{t}(n_{2},n_{1}, n_{3})$
 		 \& \\$\mbox{MACD}_{t}(n_{2},n_{1}) < \mbox{MACDS}_{t}(n_{2},n_{1}, n_{3})$\\[1mm]
		 otherwise}\\ 
	 
	 	Fast Stochastics &  \makecell[lt]{\noindent\textit{Buy} \\[1mm] \textit{Sell} \\[1mm] \textit{Hold}} & \makecell[lt]{$\mbox{Fast\%K}_{t-1}(n) \leq \mbox{Fast\%D}_{t}(n)$
		 \& $\mbox{Fast\%K}_{t}(n) > \mbox{Fast\%D}_{t}(n)$\\[1mm]
		 $\mbox{Fast\%K}_{t-1}(n) \geq \mbox{Fast\%D}_{t}(n)$ \& $\mbox{Fast\%K}_{t}(n) < \mbox{Fast\%D}_{t}(n)$\\[1mm]
			otherwise}\\
	 
	 	Slow Stochastics &  \makecell[lt]{\noindent\textit{Buy} \\[1mm] \textit{Sell} \\[1mm] \textit{Hold}} & \makecell[lt]{$\mbox{Slow\%K}_{t-1}(n) \leq \mbox{Slow\%D}_{t}(3)$ \& $\mbox{Slow\%K}_{t}(n) > \mbox{Slow\%D}_{t}(3)$\\[1mm]
		 $\mbox{Slow\%K}_{t-1}(n) \geq \mbox{Slow\%D}_{t}(3)$
		 \& $\mbox{Slow\%K}_{t}(n) < \mbox{Slow\%D}_{t}(3)$\\[1mm]
			otherwise}\\
		
	 	RSI &  \makecell[lt]{\noindent\textit{Buy} \\[1mm] \textit{Sell} \\[1mm] \textit{Hold}} & \makecell[lt]{$\mbox{RSI}_{t-1}(n) \leq 30$ \& $\mbox{RSI}_{t}(n) > 30$\\[1mm]
		$\mbox{RSI}_{t-1}(n) \geq 70$ \& $\mbox{RSI}_{t}(n) < 70$\\[1mm]
		otherwise}\\
	
		MARSI & \makecell[lt]{\noindent\textit{Buy} \\[1mm] \textit{Sell} \\[1mm] \textit{Hold}} & \makecell[lt]{$\mbox{MARSI}_{t-1}(n) \leq 30$ \& $\mbox{MARSI}_{t}(n) > 30$\\[1mm]
		$\mbox{MARSI}_{t-1}(n) \geq 70$ \& $\mbox{MARSI}_{t}(n) < 70$\\[1mm]
		otherwise}\\
	
		Bollinger & \makecell[lt]{\noindent\textit{Buy} \\[1mm] \textit{Sell} \\[1mm] \textit{Hold}} & \makecell[lt]{$P^{c}_{t-1}\geq \mbox{Boll}^{d}_{t}(n)$ \& $P^{c}_{t}\geq \mbox{Boll}^{u}_{t}(n)$\\[1mm]
		 $P^{c}_{t-1} \leq \mbox{Boll}^{d}_{t}(n)$ \& $P^{c}_{t} > \mbox{Boll}^{u}_{t}(n)$ \\[1mm]
			otherwise}\\
		
		PROC & \makecell[lt]{\noindent\textit{Buy} \\[1mm] \textit{Sell} \\[1mm] \textit{Hold}} & \makecell[lt]{$\mbox{PROC}_{t-1}(n) \leq 0$ \& $\mbox{PROC}_{t}(n) > 0$\\[1mm]
		$\mbox{PROC}_{t-1}(n) \geq 0$ \& $\mbox{PROC}_{t}(n) < 0$\\[1mm]
		 otherwise}\\
	 
	 	Williams \%R & \makecell[lt]{\noindent\textit{Buy} \\[1mm] \textit{Sell} \\[1mm] \textit{Hold}} & \makecell[lt]{$\mbox{Will}_{t-1}(n) \geq -20$ \& $\mbox{Will}_{t}(n) < -80$\\[1mm]
		 $\mbox{Will}_{t-1}(n) \leq -20$ \& $ \mbox{Will}_{t}(n) > -80$\\[1mm]
		otherwise}\\
	 
	 	SAR & \makecell[lt]{\noindent\textit{Buy} \\[1mm] \textit{Sell} \\[1mm] \textit{Hold}} & \makecell[lt]{$\mbox{SAR}_{t-1} \geq P_{t-1}^{c}$ \ \& \ $\mbox{SAR}_{t} < P_{t}^{c}$\\[1mm]
		$\mbox{SAR}_{t-1} \leq P_{t-1}^{c}$ \ \& \ $\mbox{SAR}_{t} > P_{t}^{c}$\\[1mm]
		otherwise}\\
	 	
		\botrule
\end{longtable}
\end{small}
}

\section{Trend Following and Contrarian Mean Reversion Strategies}\label{ssec:opsrules}
\subsection{Zero-Cost BCRP}
Zero-cost BCRP is the zero-cost long/short version of the BCRP strategy and is a trend following algorithm in that long positions are taken in stocks during upward trends while short positions are taken during downward trends. The idea is to first find the portfolio controls that maximise the expected utility of wealth using all in-sample price relatives according to a given constant level of risk aversion. The resulting portfolio equation is what is known to be the Mutual Fund Separation Theorem (\cite{lee}). The second set of portfolio controls in the mutual fund separation theorem (active controls) is what we will use as the set of controls for the zero-cost BCRP strategy and are given by:
\begin{align}\label{eq:OTP}
\textbf{b} = \frac{1}{\gamma} \Sigma^{-1}\bigg[\EX[\textbf{R}] - \boldsymbol{1}\frac{\boldsymbol{1}^\intercal \Sigma^{-1} \EX[\textbf{R}]}{\boldsymbol{1}^\intercal \Sigma^{-1}\boldsymbol{1}} \bigg]
\end{align}

where $\Sigma^{-1}$ is the inverse of the covariance matrix of returns for all $m$ stocks, $\EX[\textbf{R}]$ is vector of expected returns of the stocks, $\boldsymbol{1}$ is a vector of ones of length $m$ and $\gamma$ is the risk aversion parameter. \cref{eq:OTP} is the risky Optimal Tactical Portfolio (OTP) that takes optimal risky bets given the risk aversion $\gamma$. The risk aversion is selected during each period $t$ such that the controls are unit leverage and hence $\sum_{i=1}^{m} \vert b_{i}\vert = 1$. 

The covariance matrix and expected returns for each period $t$ are computed using the set of price relatives from today back $\ell$ days (short-term look-back parameter). 

\subsection{Zero-Cost Anti-BCRP}
Zero-cost anti-BCRP (contrarian mean reverting) is exactly the same as zero-cost BCRP except that we reverse the sign of the expected returns vector such that $\EX[\textbf{R}]$ = $-\EX[\textbf{R}]$. 

\subsection{Zero-Cost Anti-Correlation}
Zero-cost anti-correlation (Z-Anticor) is a contrarian mean reverting algorithm and is an adapted version of the Anticor algorithm developed in \cite{beststock}. The first step is to extract the price relatives for the two most recent sequential windows each of length $\ell$. Let $\boldsymbol{\mu}_{2}^{\ell}$ and $\boldsymbol{\mu}_{1}^{\ell}$ denote the average log-returns of the $\ell$ price relatives in the most recent window ($\boldsymbol{x}_{t}^{t-\ell+1}$) and the price relatives in the window prior to that ($\boldsymbol{x}_{t-\ell}^{t-2\ell+1}$) respectively. Also, let the lagged covariance matrix and lagged correlation matrix be defined as follows:
\begin{align}
\Sigma^{\ell} = \frac{1}{\ell - 1}[(\boldsymbol{x}_{t-\ell}^{t-2\ell+1} - 1) - \boldsymbol{1}^{\intercal}\boldsymbol{\mu}_{1}^{\ell}]^{\intercal} [(\boldsymbol{x}_{t}^{t-\ell+1} - 1) - \boldsymbol{1}^{\intercal}\boldsymbol{\mu}_{2}^{\ell}]
\end{align}
\begin{align}
\Rho_{ij}^{\ell} = \frac{\Sigma^{\ell}_{ij}}{\sqrt{\Sigma^{\ell}_{ij}\Sigma^{\ell}_{ij}}}
\end{align}

Z-Anticor then computes the claim that each pair of stocks have on one another, denoted claim$_{i\rightarrow j}^\ell$, which is the claim of stock $j$ on stock $i$ and is the extent to which we want to shift our allocation from stock $i$ to stock $j$ (\cite{beststock}). claim$_{i\rightarrow j}^\ell$ exists and is thus non-zero if and only if $\boldsymbol{\mu}_{2} > \boldsymbol{\mu}_{1}$ and $\Rho_{ij} > 0$. The claim is then calculated as
\begin{align}
\text{claim}_{i\rightarrow j}^\ell = \Rho_{ij}^{\ell} + \text{max}(-\Rho_{ii}^{\ell},0) + \text{max}(-\Rho_{jj}^{\ell},0)
\end{align}

The adaptation we propose for long/short portfolios for the amount of transfer that take places from stock $i$ to stock $j$ is given by:
\begin{align}
\text{transfer}_{i\rightarrow j}^\ell = \frac{1}{3}\text{claim}_{i\rightarrow j}^\ell
\end{align}

Finally, we calculate the expert control for the $i^{th}$ stock in period $t+1$ as follows:
\begin{align}
\textbf{h}^{n}_{t+1}(i) = \textbf{h}^{n}_{t}(i) + \sum_{i}[\text{transfer}_{j\rightarrow i}^\ell -\text{transfer}_{i\rightarrow j}^\ell]
\end{align}

Each control is then normalised in order to ensure unit leverage on the set of controls.


\begin{thebibliography}{99}

\bibitem[\protect\citeauthoryear{Algoet and Cover}{1988}]{algoet}
Algoet, P.H. and Cover, T.M., Asymptotic optimality and asymptotic equipartition properties of logoptimum investment. {\itshape The Annals of Probability}, 1988, {\bfseries 16}, 876-898. %https://doi.org/10.1214/aop/1176991793.

\bibitem[\protect\citeauthoryear{Aronson}{2007}]{Aronson}
Aronson, D.R., {\itshape Evidence-Based Technical Analysis: Applying the Scientific Method and Statistical Inference to Trading Signals}, 2007 (John Wiley \& Sons Inc: Hoboken, New Jersey).

\bibitem[\protect\citeauthoryear{Akcoglu {\itshape et al.}}{2004}]{marginaccount}
Akcoglu, K., Drineas, P. and Kao, M-Y., Fast Universalization of Investment Strategies. {\itshape SIAM Journal on Computing}, {\bfseries 34}, 1-22.

\bibitem[\protect\citeauthoryear{Andersen {\itshape et al.}}{2001}]{Andersen}
Andersen, T., Bollerslev, T., Diebold, F.X. and Labys, P., The Distribution of Realized Exchange Rate Volatility. {\itshape Journal of the American Statistical Association}, {\bfseries 96}, 42-55.

\bibitem[\protect\citeauthoryear{Bailey {\itshape et al.}}{2014}]{bailey} 
Bailey, D.H., Borwein, J.M., de Prado, M.L. and Zhu, Q.J., Pseudo-Mathematics and Financial Charlatanism: The Effects of Backtest Overfitting on Out-of-Sample Performance. {\itshape Notices of the AMS}, 2014, {\bfseries 61}, 458-471.

\bibitem[\protect\citeauthoryear{Bailey {\itshape et al.}}{2016}]{pbo}
Bailey, D.H., Borwein, J.M., de Prado, M.L. and Zhu, Q.J., The Probability of Backtest Overfitting. {\itshape Journal of Computational Finance}, 2016, {\bfseries 20}, 39-69. %DOI:10.21314/JCF.2016.322. 

\bibitem[\protect\citeauthoryear{Bailey and L{\'o}pez de Prado}{2014}]{Bailey2}
Bailey, D.H. and L{\'o}pez de Prado, M.M., The Deflated Sharpe Ratio: correcting for selection bias, backtest overfitting and non-normality. {\itshape Journal of Portfolio Management}, 2014, {\bfseries 40}, 94-107. 

\bibitem[\protect\citeauthoryear{Black}{1986}]{black}
Black, F., Noise. {\itshape Journal of Finance}, 1986, {\bfseries 41}, 529-543.

\bibitem[\protect\citeauthoryear{Bouchaud {\itshape et al.}}{2017}]{bouchaudfactor2}
Bouchaud, J-P, Ciliberti, S., Lemperiere, Y, Majewski, A., Seager, P., Ronia, K.S., Black Was Right: Price Is Within a Factor 2 of Value, 2017. Available at SSRN: https://ssrn.com/abstract=3070850.

\bibitem[\protect\citeauthoryear{Bouchaud {\itshape et al.}}{2004}]{Bouchaud}
Bouchaud, J-P, Gefen, Y., Potters, M. and Wyart, M., Fluctuations and Response in Financial Markets: The Subtle Nature of 'Random' Price Changes, 2004. Available at SSRN: https://ssrn.com/abstract=507322.

\bibitem[\protect\citeauthoryear{Borodin}{2004}]{beststock}
Borodin, A., El-Yaniv, R., and Gogan, V., Can we learn to beat the best stock. {\itshape Journal of Artificial Intelligence Research}, 2004, {\bfseries 21} , 579-594.

\bibitem[\protect\citeauthoryear{Campbell}{2005}]{backtesingreview}
Campbell, S.D., Review of Backtesting and Backtesting Procedures. {\itshape Finance and Economics Discussion Series, Board of Governors of the Federal Reserve System}, 2005, No. 2005-21.

\bibitem[\protect\citeauthoryear{Cover}{1991}]{cover}
Cover, T.M., Universal Portfolios. {\itshape Math. Finance}, 1991, {\bfseries 1}, 1-29. 

\bibitem[\protect\citeauthoryear{Cover and Ordentlich}{1996}]{coverOrd}
Cover, T.M., and Ordentlich, E., Universal Portfolios with Side Information. {\itshape IEEE Transactions on Information Theory}, 1996, {\bfseries 42}, 348-363.

\bibitem[\protect\citeauthoryear{Clayburg}{2002}]{clayburg}
Clayburg, J.F., {\itshape Four Steps to Trading Success: Using Everyday Indicators to Achieve Extraordinary Profits}, 2002 (John Wiley \& Sons: Canada).

\bibitem[\protect\citeauthoryear{Chan}{2009}]{chan2009}
Chan, E., {\itshape Quantitative Trading: How to Build Your Own Algorithmic Trading Business}, 2009 (Wiley: Hoboken, New Jersey).

\bibitem[\protect\citeauthoryear{Chande and Kroll}{1994}]{chande1994}
Chande, T.S., and Kroll, S., {\itshape The new technical trader: boost your profit by plugging into the latest indicators}, 1994 (Wiley:
New York).

\bibitem[\protect\citeauthoryear{Creamer and Freund}{2010}]{creamer}
Creamer, G. and Freund, Y., Automated trading with boosting and expert weighting. {\itshape Quantitative Finance}, 2010, {\bfseries 10}, 401-420.% \url{http://dx.doi.org/10.1080/14697680903104113}

\bibitem[\protect\citeauthoryear{De Wit}{2013}]{UCTMphil}
De Wit, J-J.D., Statistical Arbitrage in South Africa. Minor dissertation submitted to the faculty of Commerce in partial fulfilment of the requirements for the Master’s degree in Mathematical Finance at the University of Cape Town, 2013.

\bibitem[\protect\citeauthoryear{Dochow}{2016}]{dochow}
Dochow, R., Online Algorithms for the Portfolio Selection Problem, 2016 (Springer: Wiesbaden).

\bibitem[\protect\citeauthoryear{``etfSA STeFI''}{2011}]{stefi}
etfSA, STeFI-linked Money Market Deposit, Nedbank Beta Solutions, 2011. Available online at: https://www.fundsdata.co.za/etf\_news/11\_04\_04\_etfsa\_stefi\_explan\_doc.pdf (accessed 15 March 2019).

\bibitem[\protect\citeauthoryear{Fama and Blume}{1996}]{blume}
Fama, E.F. and Blume, M.E., Filter rules and stock market trading. {\itshape Journal of Business}, 1996, {\bfseries 39}, 226-241.

\bibitem[\protect\citeauthoryear{Fama}{1970}]{EMH}
Fama, E.F., Efficient capital markets: A review of theory and empirical work. {\itshape Journal of Finance}, 1970, {\bfseries 25}, 383-417.

\bibitem[\protect\citeauthoryear{Farmer}{2000}]{Farmerpredprey}
Farmer, J.D., A Simple Model for the Nonequilibrium Dynamics and Evolution of a Finacnial Market. {\itshape International Journal of Theoretical and Applied Finance}, 2000, {\bfseries 3}, 425-441.

\bibitem[\protect\citeauthoryear{Fayek {\itshape et al.}}{2013}]{Marsi}
Fayek, M.B., El-Boghdadi, H.M. and Omran, S.M., Multi-objective Optimization of Technical Stock Market Indicators using GAs. {\itshape International Journal of Computer Applications}, 2013, {\bfseries 68}, 41-48. 

\bibitem[\protect\citeauthoryear{Gy{\"o}rfi {\itshape et al.}}{2012}]{gyorfibook}
Gy{\"o}rfi, L., Ottucs{\'a}k, G. and Walk, H., {\itshape Machine Learning for Financial Engineering}, 2012 (Imperial College Press: London).
	
\bibitem[\protect\citeauthoryear{Gy{\"o}rfi {\itshape et al.}}{2008}]{gyorfi} 
Gy{\"o}rfi, L., Udina, F. and Walk, H., Nonparametric nearest neighbor based empirical portfolio selection strategies. {\itshape Statistics and Decisions}, 2008 {\bfseries 26}, 145-157.

\bibitem[\protect\citeauthoryear{Gatheral}{2010}]{gatheral} 
Gatheral, J., No-dynamic arbitrage and market Impact. {\itshape Quantitative Finance}, 2010, {\bfseries 10}, 749-759.

\bibitem[\protect\citeauthoryear{Gebbie}{2010}]{techbox}
Gebbie, T., MATLAB Technical Analysis toolbox (``technical''), 2010, https://github.com/timgebbie/MATLAB-quant-finance/technical, http://dx.doi.org/10.25375/uct.11302403

\bibitem[\protect\citeauthoryear{Hogan {\itshape et al.}}{2004}]{hogan} 
Hogan, S., Jarrow, R., Teo, M. and Warachka, M., Testing market efficiency using statistical arbitrage with applications to momentum and value strategies. {\itshape Journal of Financial Economics}, 2004, {\bfseries 73}, 525-565.

\bibitem[\protect\citeauthoryear{Harvey {\itshape et al.}}{2016}]{mike}
Harvey, M., Hendricks, D., Gebbie, T. and Wilcox, D., Deviations in expected price impact for small transaction volumes under fee restructuring. {\itshape Physica A: Statistical Mechanics and it's Applications}, 2017, {\bfseries 471}, 416-426. %\url{https://doi.org/10.1016/j.physa.2016.11.042.}

\bibitem[\protect\citeauthoryear{Hou {\itshape et al.}}{2017}]{phacking}
Hou, K., Xue, C. and Zhang, L., Replicating Anomalies. {\itshape NBER Working Paper Series}, 2017, No. 23394. %\url{http://www.nber.org/papers/w23394}

\bibitem[\protect\citeauthoryear{Huberman and Stanzl}{2005}]{huberman}
Huberman, G. and Stanzl, W., Optimal Liquidity Trading. {\itshape Springer: Review of Finance}, 2005, {\bfseries 9}, 165-200. %DOI:10.1007/s10679-005-7591-5. 

\bibitem[\protect\citeauthoryear{``Ichimoku 101''}{n.d.}]{ichimoku101}
Ichimoku 101, IchimokuTrade.com, n.d.. https://www.ichimokutrade.com/articles/Ichimoku\_Ebook.pdf (accessed 28 July 2018).

\bibitem[\protect\citeauthoryear{Jarrow {\itshape et al.}}{2012}]{jarrow}
Jarrow, R., Teob, M., Tse, Y-K and Warachka, M., An improved test for statistical arbitrage. {\itshape Journal of Financial Markets}, 2012, {\bfseries 15}, 47-80.

\bibitem[\protect\citeauthoryear{Johnson {\itshape et al.}}{2013}]{johnson} 
Johnson, N., Zhao, G., Hunsader, E., Qi, H., Meng, J., and Tivnan, B., Abrupt rise of new machine ecology beyond human response time. Scientific Reports, 2013, {\bfseries 3}. %(2627). %DOI: 10.1038/srep02627.

\bibitem[\protect\citeauthoryear{Karatzas and Shreve}{1991}]{karatzas1991brownian}
Karatzas, I. and Shreve, S.E., {\itshape Brownian Motion and Stochastic Calculus}, 1991 (Springer: New York). %, ISBN: 9780387976556.

\bibitem[\protect\citeauthoryear{Kelly}{1956}]{K1956}
Kelly, J.L., A new interpretation of information rate. {\itshape Bell. Syst. Tech.}, 1956, {\bfseries 35}, 917-926.

\bibitem[\protect\citeauthoryear{Kestner}{2003}]{QuantStrat}
Kestner ,L.N., {\itshape Quantitative Trading Strategies: Harnessing the Power of Quantitative Techniques to Create a Winning Trading Program}, 2003, (McGraw-Hill: New York). 

\bibitem[\protect\citeauthoryear{Lee}{2000}]{lee}
Lee, W., Theory and methodology of tactical asset allocation, 2000 (Wiley \& Sons: New Jersey).

\bibitem[\protect\citeauthoryear{Li and Hoi}{2016}]{binli}
Li, B. and Hoi, S.C., {\itshape Online Portfolio Selection: Principles and Algorithms}, 2016 (CRC Press: Boca Raton, FL). 

\bibitem[\protect\citeauthoryear{Linton}{2010}]{linton2010cloud}
Linton, D., Cloud Charts: Trading Success with the Ichimoku Technique, 2010 (Unknown Publisher).  

\bibitem[\protect\citeauthoryear{Lo {\itshape et al.}}{2000}]{lo2000}
Lo, A., Mamaysky, V. and Wang, J., Foundations of Technical Analysis: Computational Algorithms, Statistical Inference, and Empirical Implementation. {\itshape The Journal of Finance}, 2000, {\bfseries 55}, 1705-1765.

\bibitem[\protect\citeauthoryear{Lo and Hasanhodzic}{2009}]{hereticsfinance}
Lo, A.W. and Hasanhodzic, J., {\itshape The Heretics of Finance: Conversations with Leading Practitioners of Technical Analysis}, 2009 (Bloomberg Press: New York).

\bibitem[\protect\citeauthoryear{Lo}{2002}]{lo2002}	
Lo, A., The statistics of Sharpe Ratio's. {\itshape Financial Analyst Journal}, 2002, {\bfseries 58}, 36-52.

\bibitem[\protect\citeauthoryear{Loonat and Gebbie}{2018}]{loonat} 
Loonat, F. and Gebbie, T., Learning zero-cost portfolio selection with pattern matching. {\itshape PLoS ONE}, 2018, {\bfseries 13}: e0202788. %\url{https://doi.org/10.1371/journal.pone.0202788}.

\bibitem[\protect\citeauthoryear{Loonat and Gebbie}{2016}]{loonatcode}
Loonat, F. and Gebbie, T., MATLAB Pattern Matching and Learning Class (``pattern''), 2016. Available online at: https://github.com/FayyaazL/Pattern.

\bibitem[\protect\citeauthoryear{Moffit}{2017}]{MS2017}
Moffitt, S., Why Markets Are Inefficient: A Gambling ‘Theory’ of Financial Markets for Practitioners and Theorists, February 22, 2017.

\bibitem[\protect\citeauthoryear{Murphy}{2019}]{git}
Murphy, N., Learning Technical Trading, 2019. Available online at: https://github.com/NJ-Murphy/Learning-Technical-Trading.git.

\bibitem[\protect\citeauthoryear{Murphy and Gebbie}{2019}]{mendeleydata}
Murphy, N., Gebbie, T.,JSE Equity Market Transaction Data : TRTH and BLP Daily and Intraday data, 2019, Mendeley Data, v4 http://dx.doi.org/10.17632/v5466fbpf9.4

\bibitem[\protect\citeauthoryear{Park and Irwin}{2004}]{techanreview}
Park, C-H.and Irwin, S., The Profitability of Technical Analysis: A Review. {\itshape  AgMAS Project Research Reports}, 2004, No. 04.

\bibitem[\protect\citeauthoryear{Poon}{2008}]{Poon}
Poon, S-H., Volatility Estimation, 2008. Available online at: https://www.cmegroup.com/trading/fx/files/volEst.pdf (accessed 10 September 2017).
						
\bibitem[\protect\citeauthoryear{Rechenthin}{2014}]{PHDRechenthin}
Rechenthin, M.D., Machine-learning classification techniques for the analysis and prediction of high-frequency stock direction. PhD thesis, University of Iowa, 2014. %\url{http://ir.uiowa.edu/etd/4732}.

\bibitem[\protect\citeauthoryear{Samo and Hendricks}{2018}]{aseetuseful}
Samo, Y-L.K., and Hendricks, D., What Makes An Asset Useful?, 2018. Available online at: https://arxiv.org/abs/1806.08444 (accessed 01 January 2019).

\bibitem[\protect\citeauthoryear{Schulmeister}{2009}]{Schulmeister} 
Schulmeister, S., Profitability of technical stock trading: Has it moved from daily to intraday data?. {\itshape Review of Financial Economics}, 2009, {\bfseries 18}, 190-201.
							
\bibitem[\protect\citeauthoryear{``The Hidden Costs of Trading''}{n.d.}]{spread}
The Hidden Costs of Trading, n.d.. Available online at: http://pages.stern.nyu.edu/$\sim$adamodar/New\_Home\_Page/invemgmt/trading.htm (accessed 20 March 2018).
																		
\bibitem[\protect\citeauthoryear{``The Definitive Reference to the Ichimoku Kinko Hyo''}{n.d.}]{kumotrader}
The Definitive Reference to the Ichimoku Kinko Hyo, n.d.. Available online at: http://www.kumotrader.com/ichimoku\_wiki/index.php?title=Main\_Page (accessed 10 October 2018).
					
\bibitem[\protect\citeauthoryear{Wilder}{1978}]{SAR}
Wilder, J.W., New Concepts in Technical Trading Systems, 1978 (Trend Research).
																	
\end{thebibliography}
\end{document}